\documentclass[twocolumn]{autart}

\DeclareMathAlphabet{\mathpzc}{OT1}{pzc}{m}{it}

\usepackage{ifpdf}
 \ifpdf
 \else
 \fi

\usepackage{multirow}

\usepackage[noadjust]{cite}

\usepackage{autobreak}

   \usepackage[pdftex]{graphicx}

\usepackage{amsmath}
\usepackage{amssymb}
\usepackage{mathrsfs}
\usepackage{amsfonts}
\usepackage{color}
\newtheorem{proof}{Proof of Theorem}
\newtheorem{definition}{Definition}

\newtheorem{lemma}{Lemma}
\newtheorem{remark}{Remark}

\newtheorem{proposition}{Proposition}

\newtheorem{procedure}{Procedure}
\usepackage{algpseudocode}

\usepackage{array}

\usepackage{cases}

\usepackage{stfloats}
\usepackage{url}
\begin{document}
\begin{frontmatter}
\title{Stability of Multi-Dimensional Switched Systems with an Application to Open Multi-Agent Systems \thanksref{footnoteinfo}}
\thanks[footnoteinfo]{This paper was not presented at any IFAC meeting.
(Corresponding authors: Y. Tang and F. Qian).}

\author[Paestum]{Mengqi Xue}\ead{x\_starter@hotmail.com},
\author[Paestum]{Yang Tang} \ead{tangtany@gmail.com},
\author[Rome]{Wei Ren}    \ead{ren@ece.ucr.edu},
\author[Paestum]{Feng Qian}    \ead{fqian@ecust.edu.cn}

\address[Paestum]{Key Laboratory of Smart
Manufacturing in Energy Chemical Process,
Ministry of Education, East China University of Science and Technology, Shanghai 200237, P.R. China.}
\address[Rome]{Department of Electrical and Computer Engineering,
University of California, Riverside 92521, USA.}

\date{}

\begin{abstract}
A multi-dimensional switched system or multi-mode multi-dimensional ($M^3D$) system extends the classic switched system by allowing different subsystem dimensions. The stability problem of the $M^3D$ system, whose state transitions at switching instants can be discontinuous due to the dimension-varying feature, is studied. The discontinuous state transition is formulated by an affine map that captures both the dimension variations and the state impulses, with no extra constraint imposed. In the presence of unstable subsystems, the general criteria featuring a series of Lyapunov-like conditions for the practical and asymptotic stability properties of the $M^3D$ system are provided under the proposed slow/fast transition-dependent average dwell time framework. Then, by considering linear subsystems, we propose a class of parametric multiple Lyapunov functions to verify the obtained Lyapunov-like stability conditions and explicitly reveal a connection between the practical/asymptotic stability property and the non-vanishing/vanishing property of the impulsive effects in the state transition process. Further, the obtained stability results for the $M^3D$ system are applied to the consensus problem of the open multi-agent system (MAS), whose network topology can be switching and size-varying due to the migrations of agents. It shows that through a proper transformation, the seeking of the (practical) consensus performance of the open MAS with disconnected digraphs boils down to that of the (practical) stability property of an $M^3D$ system with unstable subsystems.
\end{abstract}


\end{frontmatter}

\noindent\textbf{\scriptsize{Summary of Main Acronyms}}
\begin{table}[h]
\scriptsize
{
  \begin{tabular}{lll}
    \hline
    \textbf{Acronym} &&\textbf{Meaning} \\
    \hline
   $M^3D$ && Multi-Mode Multi-Dimensional\\
    MAS && Multi-Agent System\\
    ADT && Average Dwell Time\\
    MDADT && Mode-Dependent Average Dwell Time\\
    TDADT && Transition-Dependent Average Dwell Time\\
    MLF && Multiple Lyapunov Function\\
    GUPS&& Global Uniform Practical Stability\\
    GUAS&& Global Uniform Asymptotic Stability\\
   \hline
    \end{tabular}
    }
\end{table}
\section{Introduction}\label{section_1}
As an important branch of hybrid systems, switched systems \cite{RN1170} have received much attention over the last few decades for their simplicity and effectiveness in modeling  systems with both continuous and discontinuous dynamics, see e.g., \cite{RN268,hespanha1999stability,RN179,878825,hespanha2004uniform,RN258}, and some recent works \cite{RN1215,RN1286,RN1188,RN2048,RN1966,RN677,zhao2017new}. In most of these existing works on switched systems, a common and conventional setting is that all the subsystems (switching modes) share the same state dimension. Such a setting renders the switched system an invariant and individual state-space structure, which enables one to analyze the state evolution of the switched system in a similar fashion to traditional non-switched systems. Despite providing such a decent property, the setting of the same state dimension is somehow ideal as it may not precisely reflect the true picture of a practical system which works in different modes. {For example, a fixed-wing aircraft can undergo several transitions between the cruise and the glide phases during multiple flights. Considering these transitions instantaneous, the aircraft can then be deemed a switched system with two switching modes (corresponding to the airborne dynamics in the cruise and the ground dynamics in the glide phases, respectively). Meanwhile, the aircraft can exhibit different degrees of freedom (DOF) when airborne (e.g., 6 DOF \cite{cook2012flight}) and on the ground (e.g., 5 DOF with the oleo strut applied to the landing gear \cite{kruger1997aircraft}). If one tries to completely but not redundantly describe the motions of the aircraft in these two different phases using state-space models, then the required numbers of state variables would be different accordingly. This clearly does not satisfy the same-dimension setting and instead leads to a non-canonical  switched system that has multiple different subsystem dimensions.
Formally, one can call such a kind of switched systems
the multi-dimensional switched system.

To date, the studies on multi-dimensional switched systems remain in a minority. One of the pioneering explorations was made in \cite{verriest2006multi}, where ``multi-mode multi-dimensional ($M^3D$) system'' ({the term ``$M^3D$ system'' will also be used to denote ``multi-dimensional switched system'' in this work}) was first used to indicate switched systems with different subsystem dimensions. The authors later introduced the concept of pseudo-continuity in \cite[Definition 1]{RN5632} to the $M^3D$ system, such that its state trajectory can be meaningfully studied. However, such a property prohibits the situation where transitions start from a higher dimensional subsystem to a lower one and then back to a higher one, in order to avoid a possible loss of state information \cite{RN5632}. {This consequently makes the pseudo-continuous $M^3D$ model less universal}.
{The authors in \cite{RN1087} studied the stability of a set of switched linear systems which may share different state spaces. By letting the state trajectories be concatenated via the so-called gluing conditions, stability conditions in terms of linear matrix inequalities were obtained. Note that the proposed gluing conditions exclude the case where the state impulses do not vanish, which enables the seeking of the asymptotic stability but cannot fully cover the state transition situations at switching instants of an $M^3D$ system. In \cite{RN3982}, the time-variant frequency response function was employed to characterize hybrid systems with unknown complex structure that potentially implies different subsystem dimensions, and an estimation algorithm for the function was proposed based on the input and output information.}
  Generally speaking, these works have sparked the studies on the $M^3D$ system and yielded some enlightening results, though, the system models involved still lack some universality, especially for the state transitions at switching instants. Moreover, how to seek the stability of $M^3D$ systems by time-dependent switchings, especially in the presence of unstable subsystems, also deserves to be further investigated. {It is notable that there have been some endeavors on the stability problems of switched systems with unstable subsystems during the recent few years, e.g., \cite{RN1159,878825,RN1188,zhao2017new,RN1911}. Among them, the method of using stable dynamics to compensate the unstable dynamics \cite{RN1159,878825,zhao2017new} and the techniques featuring the use of the Lyapunov functions with time-dependent parameters (e.g., the discretized Lyapunov function \cite{RN1188}, and the quasi-time-dependent function \cite{RN1911}) are commonly employed to ensure the stability of the whole switched system. However, these approaches for classic switched systems may not be directly applied to the $M^3D$ system due to its more complicated dimension-varying structure. Considering that the number of relevant results is also limited, this thus motivates the corresponding part of this work.}

On the other hand, the switched system has long been linked with the multi-agent system (MAS) owing to a series of early discussions on switching networks (topologies) \cite{1205192,RN154,RN1281}.
In general, the commonly considered switching topology exhibits instantaneous variations in network connections, whereas the network scale is usually considered to be fixed, see e.g., \cite{RN154,RN1281,RN1907,RN648,RN988,RN921}. However, such a setting could not hold for the MAS networks whose nodes (agents) show dynamic flowing (or migration, e.g., arrivals or departures of agents) behaviors. For example, some agents may go offline at certain instants due to faults and go back online after certain periods with faults fixed. This will cause the scale of the network to intermittently vary. The MAS with such variations in the network scale (potentially in connections) is termed the open MAS \cite{demazeau1996populations,hendrickx2016open}.

In contrast to its early popularity in the computer community \cite{demazeau1996populations}, the open MAS has not received much attention until recently in the control community.
An initial effort on the open MAS was made by \cite{hendrickx2016open}, {in which the authors studied the consensus problem under the gossip algorithm that randomly selects a pair of agents at a certain time instant and then calculates their averages to update their states}. To deal with the scale variations of the network, the ``scale-independent'' quantities were considered therein as the metrics for consensus errors. The agent migration behaviors were considered to be deterministic, which was also assumed in \cite{abdelrahim2017max} for the max consensus problem. The result in \cite{hendrickx2016open} was then extended by \cite{hendrickx2017open} from the deterministic migration case to the random case. Note that all the aforementioned works had established their results on an implicit assumption of a completely connected communication graph, which indicates each pair of agents remain connected.
Further, the proportional dynamic consensus problem is studied for the open MAS by \cite{franceschelli2018proportional}, in which the authors introduced an open distance function to illustrate the consensus error and proposed a formal stability definition for the error trajectories. However, the results were also based on the assumption that the considered directed graph is strongly connected every time.
In general, from the aforementioned results on the open MAS, it can be seen that they were obtained either by minimizing the impact of the size-varying property of the topology
\cite{hendrickx2016open,hendrickx2017open} or by specifying some strong graph connectivity conditions  \cite{abdelrahim2017max,franceschelli2018proportional}. Results on consensus problems of the open MAS that take the size-varying topology into account while rely on more relaxed graph connectivity settings are still lacking.
On the other hand, the naturally switching and size-varying feature of the open MAS network has shown a close relation to the feature of the $M^3D$ system under discussion. In light of the well-known applications of conventional switched systems to the MASs with switching topologies, it is then of interest to seek for a possible application of the $M^3D$ system to the open MAS.

Motivated by the above, this work will focus on the $M^3D$ system as well as its application to the open MAS. The main contributions are highlighted as follows:
{
\begin{enumerate}
 \item {The $M^3D$ system, which extends classic switched systems by allowing different subsystem dimensions, is studied. The state transition at each switching instant is formulated by an affine map to characterize the potential dimension variation and non-vanishing impulse. Compared with other existing works like \cite{RN5632}, no extra constraint needs to be imposed on the state transition process.}
      \item {In the presence of unstable subsystems and potential non-vanishing impulses in the state transition process, the criteria for the practical stability of the $M^3D$ system, which feature new dwell-time concepts and Lyapunov-like conditions that extend some existing results as in \cite{RN1286,zhao2017new}, are provided. For the linear subsystem case, these stability criteria are verified by a new class of parametric multiple Lyapunov functions.}
          \item  {The $M^3D$ system is applied to address the consensus problem of the open MAS, whose network structure is switching and size-varying due to the agent migration behaviors. Compared with existing works like \cite{hendrickx2016open,abdelrahim2017max} that entail strong assumptions on graphs, we allow for open MASs with disconnected digraphs. By revealing the correspondence between the connectivity of the size-varying switching digraph and the stability of the subsystem, the consensus conditions for the open MAS with disconnected digraphs are established based on the stability result obtained for the $M^3D$ system with unstable subsystems.
              }
\end{enumerate}
}

The rest of the paper is organized as follows: Section \ref{section_2} provides the system formulation and preliminaries; Section \ref{section_3} presents the results on the stability of $M^3D$ systems;  an application of the $M^3D$ system to the open MAS is presented in Section \ref{section_4}, where a simulation result is also included; Section \ref{section_5} gives the conclusion and some prospects.

The notations used in this work are summarized as follows: $\textbf{1}_{n}$ denotes an $n\times 1$ vector that is fully composed of ones;
$\mathbb{N}$ and $\mathbb{N}_{\geq 0}$ denote the sets of positive and non-negative integers, respectively; $\mathbb{R}$ and $\mathbb{R}_{\geq 0}$ denote the sets of real and non-negative real  numbers, respectively;
$\mathbb{B}^{m\times n}$ denotes the set of $m\times n$ 0-1 matrices;
$\mathbb{R}^n$ and $\mathbb{R}^{m\times n}$ denote the sets of $n\times 1$ real vectors and $m\times n$ real matrices, respectively; $\mathbb{C}^n$ and $\mathbb{C}^{m\times n}$ denote the sets of $n\times 1$ complex vectors and $m\times n$ complex matrices, respectively;
the $n\times n$ identity matrix is denoted by $I_n$; the Kronecker product of matrices $A$ and $B$ is denoted by $A\otimes B$; $\lambda(R)$ denotes the spectrum of a square matrix $R\in\mathbb{C}^{n\times n}$ and $\lambda_i(R)$ denotes the $i$-th eigenvalue of $R$, $i\in\{1,...,n\}$; $|\mathcal{S}|$ denotes the cardinality of
a set $\mathcal{S}$; $\|...\|$ denotes the induced 2-norm of a matrix or the 2-norm of a vector; $\mathrm{Re}(...)$ denotes the real part of a complex number; $P>0$ denotes a real positive definite matrix; for a real symmetric matrix $P$, $\lambda_{\max}(P)$ and $\lambda_{\min}(P)$ denote its maximum and minimum eigenvalues, respectively.
\section{System formulation and preliminaries}\label{section_2}
In this section, we will give the mathematical description of the considered system. Meanwhile, some related concepts will also be provided as preliminaries.

\subsection{System dynamics}\label{section_2_1}
 Given a Zeno-free (finite {number of} discontinuities in any finite time interval) switching signal $\sigma(t)$, $\sigma: \mathbb{R}_{\geq 0}\rightarrow \mathcal{P}$, where $\mathcal{P}=\{1,2,...,s\}$ is the {index}-set of all $s$ subsystems, an $M^3D$ system with general nonlinear subsystem dynamics or the nonlinear $M^3D$ system is formulated as:
\begin{align}
\label{MDSS}
\dot{x}_{\sigma(t)}(t)= f_{\sigma(t)}(x_{\sigma(t)}(t)),
\end{align}
where $x_{\sigma(t)}(t)=[{x}_{\sigma(t),1}(t),{x}_{\sigma(t),2}(t),...,{x}_{\sigma(t),n_{\sigma(t)}}(t)]^T\in \mathbb{R}^{n_{\sigma(t)}}$ is the state vector, ${x}_{\sigma(t),i}(t)\in \mathbb{R}$ is the $i$-th component of $x_{\sigma(t)}(t)$, $i\in\{1,...,n_{\sigma(t)}\}$ {with $n_{\phi}<+\infty$ for any $\phi\in\mathcal{P}$},
$f_{\phi}:\mathbb{R}^{n_{\phi}}\rightarrow \mathbb{R}^{n_{\phi}}$ is locally Lipschitz w.r.t. $x_{\phi}(t)$ and $f_{\phi}(0)=0$ for each $\phi\in\mathcal{P}$. {The stability of a subsystem $\phi$ is defined about its equilibrium $0$}. The switching signal $\sigma(t)$ is a right-continuous piecewise {constant} function, i.e., $\sigma(t_k)=\sigma(t_k^+)$, where $t_k$, $k\in\mathbb{N}$ denotes the $k$-th discontinuous (switching) instant of $\sigma(t)$. Besides, other functions of $t$ in this work are also assumed to be right continuous.
{For \eqref{MDSS},
denote by $\mathcal{P}_s$ the set of indices of all the subsystems with asymptotically stable equilibria, and denote by $\mathcal{P}_u$ the set of indices of all the subsystems with unstable or marginally stable equilibria. One then has that $\mathcal{P}_u\bigcap\mathcal{P}_s=\emptyset$ and $\mathcal{P}_u\bigcup\mathcal{P}_s=\mathcal{P}$.}
Particularly, the $M^3D$ system with general linear subsystem dynamics or briefly the linear $M^3D$ system can be formulated as the following closed-loop model with a state feedback control:
\begin{align}
\label{MDSS_linear_closed_loop}
\dot{x}_{\sigma(t)}(t)= (A_{\sigma(t)}+B_{\sigma(t)}K_{\sigma(t)})x_{\sigma(t)}(t),
\end{align}
where $A_{\sigma(t)}\in \mathbb{R}^{n_{\sigma(t)} \times n_{\sigma(t)}}$, $B_{\sigma(t)}\in \mathbb{R}^{n_{\sigma(t)} \times r}$, $K_{\sigma(t)}\in\mathbb{R}^{r\times n_{\sigma(t)}}$, $r$ is a certain positive integer.
{It can be seen from \eqref{MDSS} that the system state can exhibit different dimensions with the evolution of the switching signal $\sigma(t)$, which implies a dimension-varying property of the $M^3D$ system. This thus makes the $M^3D$ system contain the classic switched system as a special case (by setting $n_{\sigma(t)}$ constant for all $t$).
Note that by $\mathbb{R}^{n_{\sigma(t)}}$, we do not mean that the state space itself actually evolves with time, but that the system state evolves through state spaces with different structures under the evolution of $\sigma(t)$.}

\begin{remark}\upshape
The $M^3D$ systems extend classic switched systems by allowing different subsystem dimensions. Meanwhile, the dimension-varying property also potentially complicates the corresponding analysis. Given this, one may prefer to circumvent a direct analysis of an $M^3D$ system by converting it into a classic switched system. Intuitively, two methods are adoptable to implement such a conversion. The first is to apply the model reduction technique to the $M^3D$ system, such that the reduced subsystem models have the same dimension {(an opposite situation where the model reduction technique converts classic switched systems into $M^3D$ systems was also suggested in \cite{verriest2006multi})}. However, this method may come at the cost of losing part of the state information. The second is to insert extra components into each subsystem state such that the resultant subsystems have the equal dimension \cite{wang2008stability}. Compared with the first one, the second method does not inflict a loss of information, though, the introduced components would potentially cause an unnecessary increase of the computation burden. Besides, the second method is also ineffective in the case where the highest allowed dimension is unknown or unfixed. Thus, it is always meaningful for one to consider a direct analysis of the $M^3D$ system.
\end{remark}

The equation \eqref{MDSS} characterizes the dynamics of an $M^3D$ system in each non-switching period. However,  it alone is insufficient to determine a complete system behavior. One still needs the switching-time behaviors to bridge all these non-switching dynamics together. This thus leads to the following subsection.

\subsection{State transitions at switching instants}\label{section_2_2}
It is well understood that a typical switched system as studied in \cite{RN1170} always admits a piecewise differentiable state trajectory with a continuous state transition at each switching instant, i.e., $x(t_k^+)=x(t_k^-)$, $k\in\mathbb{N}$. However, this is in general not the case for an $M^3D$ system, since the dimension variation that takes place at any switching instant will surely render the system a discontinuous state transition in the meantime. Moreover, a discontinuous state transition process can also be brought by the traditional state jump/impulse as considered for impulsive systems \cite{RN1966}. Taking both the aforementioned factors into account, we can thus formulate the state transition of the $M^3D$ system \eqref{MDSS} at each switching instant $t_k$, $k\in\mathbb{N}$ as follows:
\begin{align}\label{Mode_transform_var}
x_{\sigma(t_k^+)}(t_{k}^+)=&\Xi_{\sigma(t_k^+),\sigma(t_k^-)}x_{\sigma(t^-_k)}(t_{k}^-)+\Phi_k,
\end{align}
where  $\Xi_{\sigma(t_k^+),\sigma(t_k^-)}\in\mathbb{B}^{n_{\sigma(t_k^+)}\times n_{\sigma(t_k^-)}}$ and $\Phi_{k}\in\mathbb{R}^{n_{\sigma(t_k^+)}}$.
$\Xi_{\sigma(t_k^+),\sigma(t_k^-)}$ is a special 0-1 matrix indicating the dimension variation (e.g., reduction or expansion) of $x_{\sigma(t_k^-)}(t_k^-)$ at $t_k$, which is obtained by removing certain rows from (dimension reduction) or by inserting zero rows into (dimension expansion) certain positions of an identity matrix $I_{n_{\sigma(t_k^-)}}$. On the other hand, $\Phi_{k}$ is a real vector indicating the impulse occurring to the transformed state $\Xi_{\sigma(t_k^+),\sigma(t_k^-)}x_{\sigma(t_k^-)}(t_k^-)$ at $t_k$.
An illustration of the state transition process \eqref{Mode_transform_var} characterized by $\Xi_{\sigma(t_k^+),\sigma(t_k^-)}$ and $\Phi_k$ of the $M^3D$ system \eqref{MDSS} is provided in Fig. \ref{MDSS_fig} (see the caption for detailed descriptions).

\begin{figure*}[t!]
\centering
\includegraphics[width=1\textwidth]{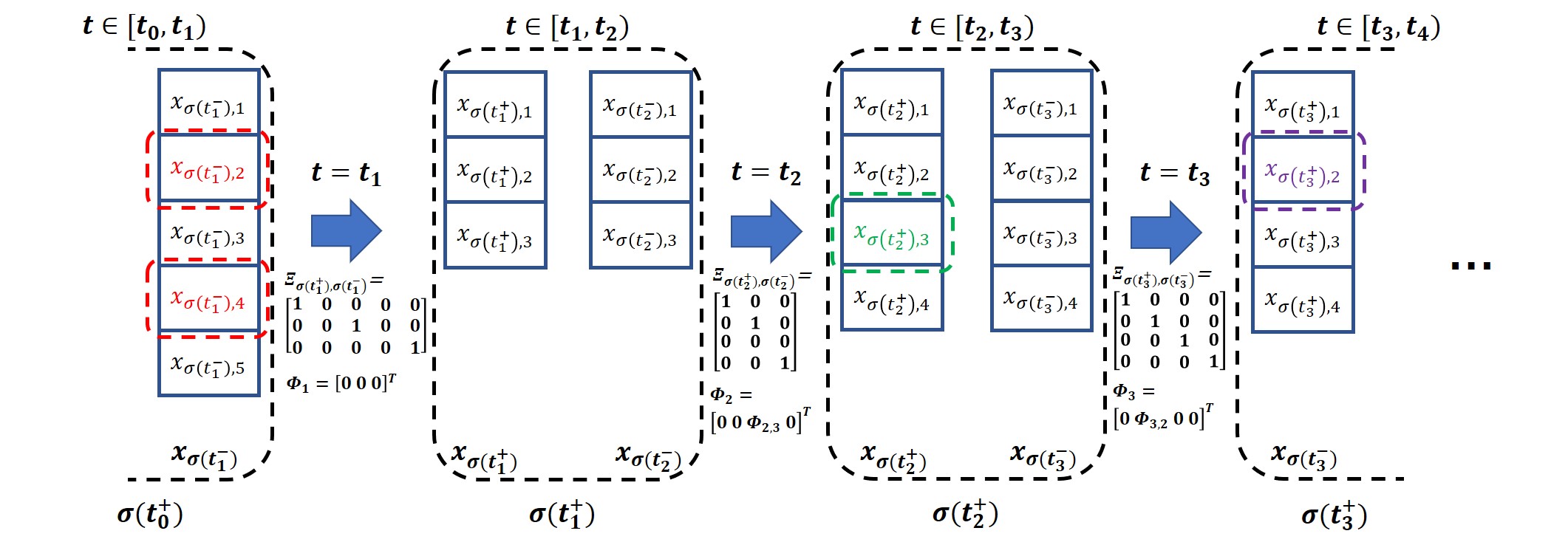}
\caption{Illustration of the state transitions at switching instants $t_1$, $t_2$, $t_3$ of an $M^3D$ system. The smaller colored dashed box enclosing a state component (in a blue solid box) denotes that the component is under dimension reduction (red) or dimension expansion (green) or state impulse (purple). The transition at $t_1$ features pure dimension reductions on $x_{\sigma(t_1^-),2}$ and $x_{\sigma(t_1^-),4}$, while other components of $x_{\sigma(t_1^-)}$ remain unchanged in value; the transition at $t_2$ features a pure dimension expansion between $x_{\sigma(t_2^-),2}$ and $x_{\sigma(t_2^-),3}$, in which the newly added component $x_{\sigma(t_2^+),3}$ is assigned the value of $\Phi_{2,3}$ while all the components of $x_{\sigma(t_2^-)}$ remain unchanged in value; the transition at $t_3$ does not exhibit any change in dimension but it features a pure state impulse brought by $\Phi_{3,2}$ to $x_{\sigma(t_3^-),2}$ that yields $x_{\sigma(t_3^+),2}$ with a different value, while other components of $x_{\sigma(t_3^-)}$ remain unchanged in value.}
\label{MDSS_fig}
\end{figure*}

\begin{remark}\upshape\label{re_1}
 {The state transition process \eqref{Mode_transform_var} is pivotal to determine a complete evolution of the $M^3D$ system \eqref{MDSS} as it defines how the system behaves at each switching instant.
 The above parameter settings for the affine expression of \eqref{Mode_transform_var} do not cause any loss of generality, since given any pair of vectors $x_{\sigma(t_k^+)}(t_k^+)\in\mathbb{R}^{n_{\sigma(t_k^+)}}$ and $x_{\sigma(t_k^-)}(t_k^-)\in\mathbb{R}^{n_{\sigma(t_k^-)}}$, each type of dimension variations between them can be fulfilled by a unique $0$-$1$ matrix $\Xi_{\sigma(t_k^+),\sigma(t_k^-)}$, and any value jump that is not the result of dimension variations can be captured by $\Phi_k$. {Note that the formulation \eqref{Mode_transform_var} does not impose any extra constraint on the state transition process compared with the ``pseudo-continuity'' required in \cite{RN5632}.} {Specifically, the case of potential ``loss of information'' when the dimension decreases and then increases that was disallowed by the ``pseudo-continuity'', can now be interpreted as the self-impulse of the state brought by the offset term $\Phi_k$.} Moreover, it is clear that the case of $\Xi_{\sigma(t_k^+),\sigma(t_k^-)}=I_{n_{\sigma(t_k^-)}}$ and $\Phi_k=0$ indicates a trivial continuous state transition at $t_k$. The case of $\Xi_{\sigma(t_k^+),\sigma(t_k^-)}=I_{n_{\sigma(t_k^-)}}$ and $\Phi_k\neq0$ indicates a pure impulse between states with the same dimension.
}
\end{remark}

\subsection{Two types of state impulses}\label{section_2_3}
The discontinuity of \eqref{Mode_transform_var} is brought by the dimension variation indicated by $\Xi_{\sigma(t_k^+),\sigma(t_k^-)}$ and/or the state impulse indicated by $\Phi_k$. In particular, it is notable that for a pure dimensional transformation at $t_k$, i.e., $x_{\sigma(t_k^+)}(t_k^+)=\Xi_{\sigma(t_k^+),\sigma(t_k^-)}x_{\sigma(t_k^-)}(t_k^-)$, the value of $x_{\sigma(t_k^+)}(t_k^+)$ always linearly depends on that of the state $x_{\sigma(t_k^-)}(t_k^-)$ through $\Xi_{\sigma(t_k^+),\sigma(t_k^-)}$.
We thus call $x_{\sigma(t_k^+)}(t_k^+)$ \textit{state-dependent} under such a pure dimensional transformation. It can also be concluded that a pure dimensional transformation always yields a state-dependent $x_{\sigma(t_k^+)}(t_k^+)$. However, when one takes the state impulse brought by $\Phi_k$ into account, then the previously defined state dependency might not necessarily hold for $x_{\sigma(t_k^+)}(t_k^+)$ since its value will also depend on that of $\Phi_k$. Given this, we consider to classify $\Phi_k$ into the following two types:

\textit{a) State-independent} $\Phi_k$: The value of $\Phi_k$ relies solely on the switching instant $t_k$ or $k$, i.e., there is no explicit relation between
$\Phi_k$ and $x_{\sigma(t_k^-)}(t_k^-)$, and satisfies
 $\|\Phi_ {k}\|\leq\bar{\Phi}$ where $\bar{\Phi}>0$ is a certain constant. This indicates $x_{\sigma(t_k^+)}(t_k^+)$ does not linearly rely on the state $x_{\sigma(t_k^-)}(t_k^-)$ as the pure dimensional transformation case. Note that in this case, $\Phi_k\not\equiv0$ and can be deemed \textit{unknown} except the upper bound $\bar{\Phi}$ of its norm.

 \textit{b) State-dependent} $\Phi_k$: The value of $\Phi_k$ relies on that of $x_{\sigma(t_k^-)}(t_k^-)$.  Specifically, we have the following explicit formulation for state-dependent $\Phi_k$:
\begin{align}\label{Phi_value}
  \Phi_{k}=\hat{\Xi}_{\sigma(t^+_k),\sigma(t_k^-)}x_{\sigma(t_k^-)}(t_k^-),
\end{align}
in which $\hat{\Xi}_{\sigma(t^+_k),\sigma(t_k^-)}\in\mathbb{R}^{n_{\sigma(t_k^+)}\times n_{\sigma(t_k^-)}}$ is a given matrix.
As a result, the state transition \eqref{Mode_transform_var} with state-dependent $\Phi_k$  in \eqref{Phi_value} can be rewritten as:
\begin{align}\label{Mode_transform_var_asymp}
x_{\sigma(t_k^+)}(t_{k}^+)=&\check{\Xi}_{\sigma(t^+_k),\sigma(t_k^-)}x_{\sigma(t^-_k)}(t_{k}^-),
\end{align}
where $\check{\Xi}_{\sigma(t^+_k),\sigma(t_k^-)}=\Xi_{\sigma(t^+_k),\sigma(t_k^-)}+\hat{\Xi}_{\sigma(t^+_k),\sigma(t_k^-)}$.
\begin{remark}\upshape
{The above two types of the impulse $\Phi_k$ reflect two evolution properties it could have.
For the state-independent $\Phi_k$, its evolution does not rely on any variable except $k$, which means that in general it will never spontaneously converge to zero regardless of the evolution of the state $x_{\sigma(t_k^-)}(t^-_k)$. We thus say that the state-independent $\Phi_k$ has a \textit{non-vanishing property}. For the state-dependent $\Phi_k$ that satisfies \eqref{Phi_value}, its evolution linearly depends on the state $x_{\sigma(t_k^-)}(t_k^-)$ through the matrix $\hat{\Xi}_{\sigma(t^+_k),\sigma(t_k^-)}$, which indicates that it would potentially converge to zero with a convergent $x_{\sigma(t_k^-)}(t_k^-)$. Accordingly, we say that the state-dependent $\Phi_k$ has a \textit{vanishing property}.
Note that such non-vanishing and vanishing properties considered for the state impulses are similar to the non-vanishing and vanishing perturbations featured in perturbed systems (see e.g., \cite[Chapter 9]{khalil1996noninear}).
 As we shall see later, these two types of $\Phi_k$ can result in different stabilities for the $M^3D$ system.}
\end{remark}

\subsection{Related concepts}\label{section_2_4}
Before proceeding, some definitions will be presented. Note in this work we consider the switching signal $\sigma(t)$ and any dynamical system on a general time interval $[t_0,t_f]$, where $t_0\geq 0$ and $t_f\in(t_0,+\infty)$ denote the initial and final times of interest, respectively. Moreover, let $N(t_0,t_f)$ denote the total number of switchings on $[t_0,t_f]$.

\begin{definition}\label{DEF_GUPS}
The $M^3D$ system \eqref{MDSS}  is said to be globally uniformly practically stable (GUPS),
if there exist a class $\mathcal{KL}$ function $\beta$ and a scalar $\epsilon\geq0$ such that for any initial state $x_{\sigma(t_0)}(t_0)$ and admissible $\sigma(t)$,
\begin{align}\label{GUPS}
\|x_{\sigma(t)}(t)\|\leq \beta(\|x_{\sigma(t_0)}(t_0)\|,t-t_0)+\epsilon, \ \forall t\geq t_0,
\end{align}
 where $\epsilon$ is called the ultimate bound of $x_{\sigma(t)}(t)$ as $t\rightarrow +\infty$. Particularly, if one has $\epsilon=0$, then \eqref{MDSS} is said to be globally uniformly asymptotically stable (GUAS).
\end{definition}

The global uniform practical stability (GUPS) defined for the $M^3D$ system \eqref{MDSS} is an extension of those defined for classic switched systems (see e.g., \cite[Definition 1]{RN2107}).
The global uniform asymptotic stability (GUAS) can be deemed a special case of the GUPS when $\epsilon=0$.

The next definition is given for a switching signal $\sigma(t)$.

\begin{definition}\label{Def_TDADT}
{Consider a switching signal $\sigma(t)$ on the interval $[t_0,t_f]$. For any $[t_k,t_f]\subseteq[t_0,t_f]$, $k\in\{0,...,N(t_0,t_f)\}$, and $t\in[t_k,t_f]$, denote the number of switchings (transitions) from $\hat{\phi}\in \mathcal{P}$ to $\phi\in\mathcal{P}$ on $[t_k,t]$ by $N_{\phi,{\hat{\phi}}}(t_k, t)$ and denote the total active periods of $\phi$ pertaining to these switchings by ${T}_{\phi,{\hat{\phi}}}(t_k, t)$. Then, for the constant $\hat{N}_{\phi,{\hat{\phi}}}\geq 0$, the scalar $\overline{\tau}_{\phi,{\hat{\phi}}}(t_k,t)$ satisfying
\begin{align}\label{TDADT_slow}
N_{\phi,{\hat{\phi}}}(t_k,t)\leq \hat{N}_{\phi,{\hat{\phi}}}+\frac{{T}_{\phi,{\hat{\phi}}}(t_k, t)}{\overline{\tau}_{\phi,{\hat{\phi}}}(t_k,t)},
\end{align}
 is called the (slow) piecewise transition-dependent average dwell time (TDADT) of the switching signal $\sigma(t)$, and in particular $\overline{\tau}_{\phi,{\hat{\phi}}}(t_0,t)$ is called the (slow) TDADT of $\sigma(t)$. Correspondingly, for the constant $\hat{N}_{\phi,{\hat{\phi}}}\leq 0$, the scalar $\underline{\tau}_{\phi,\hat{\phi}}(t_k,t)$ satisfying
\begin{align}\label{TDADT_fast}
N_{\phi,{\hat{\phi}}}(t_k, t)\geq \hat{N}_{\phi,{\hat{\phi}}}+\frac{{T}_{\phi,{\hat{\phi}}}(t_k, t)}{\underline{\tau}_{\phi,{\hat{\phi}}}(t_k,t)},
\end{align}
 is called the fast piecewise TDADT of $\sigma(t)$,  and in particular $\underline{\tau}_{\phi,{\hat{\phi}}}(t_0,t)$ is called the fast TDADT of $\sigma(t)$.}
\end{definition}

\begin{remark}\upshape
{Definition \ref{Def_TDADT} extends the existing dwell-time concepts in three aspects. First, the proposed TDADT extends the existing mode-dependent average dwell time (MDADT) \cite{RN1286} and average dwell time (ADT) \cite{hespanha1999stability} by allowing the same subsystem to have different ADTs w.r.t. different predecessors. This means a TDADT is actually defined w.r.t. the switching (transition) between a subsystem and its predecessor (hence the name ``transition-dependent''). Such a transition-dependent feature also makes the TDADT more flexible in characterizing the switching of an $M^3D$ system, given that its state transition processes between a subsystem and different predecessors can be different due to the dimension-varying property.
 Second, the concept of the fast TDADT is accordingly an extension of the fast MDADT proposed in \cite[Eq. (4)]{zhao2017new}. Opposite to the slow MDADT (\cite[Eq. (3)]{zhao2017new}), the fast MDADT imposes an upper bound instead of a lower one on the average active period of a subsystem. As was pointed out in \cite{zhao2017new}, such ``fast-switching'' property usually applies to the unstable subsystems whose active periods need to be short enough to have their destabilizing dynamics properly neutralized. Third, the piecewise TDADT is a further extension of the TDADT concept, which more detailedly defines the TDADT of a subsystem $\phi$ with the predecessor $\hat{\phi}$ on \textit{each subinterval} $[t_k,t]$, $k\in\{0,...,N(t_0,t)\}$ of an interval $[t_0,t]$ of interest. As we will see later, this concept is linked with the GUPS property of the $M^3D$ system.
 Note that the piecewise property as stated in Definition \ref{Def_TDADT} can also be introduced to the ADT or MDADT analogously.
 Also note that the piecewise TDADT brings no more conservativeness than the classic (fixed) dwell time \cite{RN268}. This is because the (fixed) dwell time requires all the active periods $[t_k,t_{k+1})$ in $[t_0,t_f]$, $k=0,...,N(t_0,t_f)-1$ to have a same bound, while the piecewise TDADT only requires the average active period of the subsystem $\phi$ with the predecessor $\hat{\phi}$ on each subinterval $[t_k,t_f]\subseteq [t_0,t_f]$, $k\in\{0,...,N(t_0,t_f)\}$ to have a same bound, which means the latter contains the former as a special case.}
\end{remark}

\section{Stability analysis of $M^3D$ systems}\label{section_3}
In this section, we are going to present one of the main results of this work concerning the stability of the $M^3D$ systems \eqref{MDSS} and \eqref{MDSS_linear_closed_loop} with the state transition process \eqref{Mode_transform_var}. In the following, assign $\phi=\sigma(t^+_k)$, $\hat{\phi}=\sigma(t^-_k)$ for a switching instant $t_k\in[t_0,t_f]$ and denote the number of switchings on $[t_k,t]$ by $N(t_k,t)$, $t\in[t_k,t_f]$, $k\in\mathbb{N}$.

\subsection{Stability criteria for general nonlinear $M^3D$ system}\label{section_3_1}
The following theorem gives the stability criteria for the general nonlinear $M^3D$ system \eqref{MDSS}.
\begin{thm}\label{Thm_1}
{Consider the $M^3D$ system \eqref{MDSS} with the switching signal $\sigma(t)$ {on $[t_0,t_f]$, $0\leq t_0<t_f<+\infty$}. If for any $\phi,\hat{\phi}\in\mathcal{P}$, there exist class $\mathcal{K}_{\infty}$ functions $\underline{\kappa}$, $\overline{\kappa}$, constants $\tilde{\gamma}_{\phi}$, $\Omega_{\phi,\hat{\phi}}>0$, $\tilde{\Theta}\geq0$, and a non-negative function $V_{\phi}(t,x_{\phi}(t)):\mathbb{R}_{\geq 0}\times \mathbb{R}^{n_{\phi}}\rightarrow \mathbb{R}_{\geq0}$, such that $\forall t\in[t_0,t_f]$,
\begin{align}\label{V_bounded}
&\underline{\kappa}(\|x_{\phi}(t)\|)\leq V_{\phi}(t,x_{\phi}(t))\leq \overline{\kappa}(\|x_{\phi}(t)\|),\\
\label{V_decay}
&\dot{V}_{\phi}(t,x_{\phi}(t))\leq\tilde{\gamma}_{\phi}V_{\phi}(t,x_{\phi}(t)),\\
\label{V_jump}
&V_{\phi}(t^+_k,x_{\phi}(t_k^+))\leq  {\Omega}_{\phi,\hat{\phi}}V_{\hat{\phi}}(t^-_k,x_{\hat{\phi}}(t_k^-))+\tilde{\Theta},
\end{align}
where $\tilde{\gamma}_{\phi}>0$, $0<\Omega_{\phi,\hat{\phi}}<1$, $\forall\phi\in\mathcal{P}_u$; $\tilde{\gamma}_{\phi}<0$, $\Omega_{\phi,\hat{\phi}}>1$, $\forall\phi\in\mathcal{P}_s$, and that $\sigma(t)$ satisfies
   \begin{align}
\label{TDADT_a}
{\overline{\tau}_{\phi,\hat{\phi}}(t_j,t_f)> -\frac{\ln\Omega_{\phi,\hat{\phi}}}{\tilde{\gamma}_{\phi}},\ j=0,...,N(t_0,t_f),}
\end{align}
for any $\phi\in\mathcal{P}_s,\hat{\phi}\in\mathcal{P}$, and
\begin{align}
\label{TDADT_b}
{
\underline{\tau}_{\phi,\hat{\phi}}(t_j,t_f)<  -\frac{\ln\Omega_{\phi,\hat{\phi}}}{\tilde{\gamma}_{\phi}},\ j=0,...,N(t_0,t_f),} \end{align}
for any $\phi\in\mathcal{P}_u,\hat{\phi}\in\mathcal{P}$, then \eqref{MDSS} is GUPS with $\epsilon={\underline{\kappa}^{-1}(\tilde{\epsilon})}$, where
$\tilde{\epsilon}=\frac{\tilde{\Theta}\tilde{c}\mathrm{e}^{-\tilde{\hat{N}}}}{1-\mathrm{e}^{\overline{\varsigma}}}$,
$\overline{\varsigma}=\max\limits_{\forall \phi,\hat{\phi}\in\mathcal{P}, j}{\varsigma}^{j}_{\phi,\hat{\phi}}$,
${\varsigma}^{j}_{\phi,\hat{\phi}}=\tau_{\phi,\hat{\phi}}(t_{j},t_f)\tilde{\gamma}_{\phi}+\ln\Omega_{\phi,\hat{\phi}}$, $\forall j\in\{1,...,N(t_0,t_f)\}$,
$\tau_{\phi,\hat{\phi}}=\overline{\tau}_{\phi,\hat{\phi}}$ for any $\phi\in\mathcal{P}_s$, $\tau_{\phi,\hat{\phi}}=\underline{\tau}_{\phi,\hat{\phi}}$ for any $\phi\in\mathcal{P}_u$,
$\tilde{c}=\mathrm{e}^{\sum_{\forall \phi,\hat{\phi}\in\mathcal{P}}\hat{N}_{\phi,\hat{\phi}}\ln\Omega_{\phi,\hat{\phi}}}$, $\tilde{\hat{N}}=\min\limits_{\forall j}\sum_{\forall \phi, \hat{\phi}\in\mathcal{P}}{\varsigma}^j_{\phi,\hat{\phi}}{\hat{N}_{\phi,\hat{\phi}}}$.
Particularly, if  $\tilde{\Theta}=0$, then \eqref{MDSS} is GUAS
 with \eqref{TDADT_a}, \eqref{TDADT_b} satisfied for $j=0$.}
\end{thm}
\begin{proof}\upshape
Throughout this proof, we use $V_{\sigma(t)}(t)$ to denote $V_{\sigma(t)}(t,x_{\sigma(t)}(t))$ for brevity. It can be obtained from \eqref{V_decay} and \eqref{V_jump} that $\forall t\in[t_k,t_{k+1})$, $k=0,...,N(t_0,t_f)$,
\begin{align}\label{proof1_4}
&V_{\sigma(t)}(t)\nonumber\\ \leq &
\mathrm{e}^{\tilde{\gamma}_{\sigma(t_{k}^+)}(t-t_{k})}\bigg(\Omega_{\sigma(t_k^+),\sigma(t^-_k)}\mathrm{e}^{\tilde{\gamma}_{\sigma(t_{k-1}^+)}(t_k-t_{k-1})}\nonumber\\ &\times V_{\sigma(t_{k-1}^+)}(t^+_{k-1})+\tilde{\Theta}\bigg)
\nonumber\\
\leq & \mathrm{e}^{\tilde{\gamma}_{\sigma(t_{k}^+)}(t-t_{k})}\bigg(\Omega_{\sigma(t_k^+),\sigma(t^-_k)}\Omega_{\sigma(t_{k-1}^+),\sigma(t^-_{k-1})}\nonumber\\&\times \mathrm{e}^{\tilde{\gamma}_{\sigma(t_{k-1}^+)}(t_k-t_{k-1})}\mathrm{e}^{\tilde{\gamma}_{\sigma(t_{k-2}^+)}(t_{k-1}-t_{k-2})} V_{\sigma(t_{k-2}^+)}(t^+_{k-2})\nonumber\\&+\Omega_{\sigma(t_k^+),\sigma(t^-_k)} \mathrm{e}^{\tilde{\gamma}_{\sigma(t_{k-1}^+)}(t_k-t_{k-1})}\tilde{\Theta}+\tilde{\Theta}\bigg)\nonumber\\
&\ldots\nonumber\\
\leq &
\mathrm{e}^{\sum\limits_{{j}=1}^{k-1}\left(\ln\Omega_{\sigma(t_{k-{j}}^+),\sigma(t^-_{k-{j}})}+\tilde{\gamma}_{\sigma(t_{k-{j}}^+)}(t_{k-{j}+1}-t_{k-{j}})\right)} \nonumber\\&\times \mathrm{e}^{\ln\Omega_{\sigma(t_{k}^+),\sigma(t^-_{k})}+\tilde{\gamma}_{\sigma(t_k^+)}(t-t_k)}\mathrm{e}^{\tilde{\gamma}_{\sigma(t_{0}^+)}(t_{1}-t_{0})} V_{\sigma(t_0^+)}(t_0^+)\nonumber\\&+\sum\limits_{\nu=0}^{k-1}{\mathrm{e}^{\sum\limits_{{j}=1}^{\nu}\big(\ln\Omega_{\sigma(t_{k-{j}+1}^+),\sigma(t^-_{k-{j}+1})}+\tilde{\gamma}_{\sigma(t_{k-{j}}^+)}(t_{k-{j}+1}-t_{k-{j}})\big)}}\nonumber\\&\times \mathrm{e}^{\tilde{\gamma}_{\sigma(t_k^+)}(t-t_k)}\tilde{\Theta}.
\end{align}
In \eqref{proof1_4}, grouping the terms of the same $\phi,\hat{\phi}$ together, $\forall\phi,\hat{\phi}\in \mathcal{P}$, $\phi\neq\hat{\phi}$, and applying respectively \eqref{TDADT_slow} and \eqref{TDADT_fast} to $\phi\in\mathcal{P}_s$ and $\phi\in\mathcal{P}_u$ for $k=0$, one has:
\begin{align}\label{proof1_5}
V_{\sigma(t)}(t)\leq  & \mathrm{e}^{\sum\limits_{\forall \phi,\hat{\phi}\in\mathcal{P}}^{}(\tilde{\gamma}_{\phi}+\frac{\ln\Omega_{\phi,\hat{\phi}}}{\tau_{\phi,\hat{\phi}}(t_0,t)})T_{\phi,\hat{\phi}}(t_0,t)+\hat{N}_{\phi,\hat{\phi}}\ln\Omega_{\phi,\hat{\phi}}}\nonumber\\&\times V_{\sigma(t_0^+)}(t^+_0)
+\tilde{\Lambda}_k{\tilde{\Theta}},
\end{align}
where $\tau_{\phi,\hat{\phi}}=\overline{\tau}_{\phi,\hat{\phi}}$ for any $\phi\in\mathcal{P}_s$, $\tau_{\phi,\hat{\phi}}=\underline{\tau}_{\phi,\hat{\phi}}$ for any $\phi\in\mathcal{P}_u$, and
\begin{align}
\label{c1}
\tilde{\Lambda}_k=\sum\limits_{{j}=1}^{k}e^{\sum\limits_{\forall \phi,\hat{\phi}\in\mathcal{P}}^{}\left(\tilde{\gamma}_{\phi}T_{\phi,\hat{\phi}}(t_{k-{j}+1},t)+\ln\Omega_{\phi,\hat{\phi}}^{N_{\phi,\hat{\phi}}(t_{k-{j}+1},t)}\right)}.
\end{align}
Note that in this work we always rule out the trivial case of finite number of switchings in infinite time interval. This implies $k\rightarrow +\infty$ as $t\rightarrow +\infty$ in \eqref{c1}, which consequently makes $\tilde{\Lambda}_k$ a non-negative infinite series as $t\rightarrow +\infty$. Further, denoting ${\varsigma}^{j}_{\phi,\hat{\phi}}=\tilde{\gamma}_{\phi}{\tau_{\phi,\hat{\phi}}(t_{j},t)}+{\ln\Omega_{\phi,\hat{\phi}}}$, it follows from \eqref{TDADT_a} and \eqref{TDADT_b} that ${\varsigma}^{j}_{\phi,\hat{\phi}}< 0$, $\forall \phi, \hat{\phi}\in\mathcal{P}$, $j\in\{1,...,N(t_0,t)\}$.
Then, by respectively applying \eqref{TDADT_slow} and \eqref{TDADT_fast} to $\phi\in\mathcal{P}_s$ and $\phi\in\mathcal{P}_u$ in \eqref{c1},  one can obtain that
\begin{align}\label{proof1_6}
\tilde{\Lambda}_k
\leq&
\sum\limits_{{j}=1}^{k}\tilde{c}\mathrm{e}^{\sum\limits_{\forall \phi,\hat{\phi}\in\mathcal{P}}^{}{\varsigma}^{{k-j+1}}_{\phi,\hat{\phi}}(N_{\phi,\hat{\phi}}(t_{k-j+1},t)-\hat{N}_{\phi,\hat{\phi}})}\nonumber\\
\leq &\sum\limits_{{j}=1}^{k}\tilde{c}\mathrm{e}^{\overline{\varsigma}N(t_{k-{j}+1},t)-\tilde{\hat{N}}}\nonumber\\
\leq& \sum\limits_{{j}=1}^{k}\tilde{c}\mathrm{e}^{\overline{\varsigma}({j}-1)-\tilde{\hat{N}}}=\frac{\tilde{c}\mathrm{e}^{-\tilde{\hat{N}}}(1-\mathrm{e}^{\overline{\varsigma}k})}{1-\mathrm{e}^{\overline{\varsigma}}},
\end{align}
where $\overline{\varsigma}=\max\limits_{\forall \phi,\hat{\phi}\in\mathcal{P}, j}{\varsigma}^{j}_{\phi,\hat{\phi}}$,
$\tilde{c}=\mathrm{e}^{\sum_{\forall \phi,\hat{\phi}\in\mathcal{P}}\hat{N}_{\phi,\hat{\phi}}\ln\Omega_{\phi,\hat{\phi}}}$, $\tilde{\hat{N}}=\min\limits_{\forall j}\sum_{\forall \phi, \hat{\phi}\in\mathcal{P}}{\varsigma}^j_{\phi,\hat{\phi}}{\hat{N}_{\phi,\hat{\phi}}}$. It then follows from \eqref{proof1_6} that  $\limsup\limits_{k\rightarrow+\infty}\tilde{\Lambda}_k\leq \frac{\tilde{c}\mathrm{e}^{-\tilde{\hat{N}}}}{1-\mathrm{e}^{\overline{\varsigma}}}$, which along with
{\eqref{proof1_5} implies that $\limsup\limits_{t\rightarrow +\infty} V_{\sigma(t)}(t)\leq \tilde{\epsilon}$, where $\tilde{\epsilon}=\frac{\tilde{\Theta}\tilde{c}\mathrm{e}^{-\tilde{\hat{N}}}}{1-\mathrm{e}^{\overline{\varsigma}}}\geq 0$, and the first addend of \eqref{proof1_5} becomes a class $\mathcal{KL}$ function of $V_{\sigma(t_0^+)}(t_0^+)$, $t$. By \eqref{V_bounded} and Definition \ref{DEF_GUPS}, one further concludes that $\limsup\limits_{t\rightarrow
 +\infty}\|x_{\sigma(t)}(t)\|\leq \underline{\kappa}^{-1}(\tilde{\epsilon})$, and the $M^3D$ system \eqref{MDSS} is GUPS with $\epsilon=\underline{\kappa}^{-1}(\tilde{\epsilon})$. In particular, if $\tilde{\Theta}=0$,
 then it follows from \eqref{proof1_5} that with \eqref{TDADT_a} and \eqref{TDADT_b} satisfied for $j=0$,
$\lim\limits_{t\rightarrow +\infty}V_{\sigma(t)}(t)=0$. By \eqref{V_bounded}, this indicates $\lim\limits_{t\rightarrow
 +\infty}\|x_{\sigma(t)}(t)\|= 0$, which implies that \eqref{MDSS} is GUAS.$\blacksquare$}
\end{proof}
\begin{remark}\upshape
{
Theorem \ref{Thm_1} provides the stability criteria in terms of a Lyapunov-like or multiple Lyapunov function (MLF) $V_{\sigma(t)}$ for the $M^3D$ system \eqref{MDSS} under TDADT switchings. The conditions \eqref{TDADT_a} and \eqref{TDADT_b} impose constraints on $\sigma(t)$ in the sense of slow and fast switchings, respectively. That \eqref{TDADT_a} and \eqref{TDADT_b} are satisfied for $j>0$ implies that the system should perform piecewise TDADT switchings for the GUPS property (note that a tightest $\tilde{\epsilon}$ among different choices of $\hat{N}_{\phi,\hat{\phi}}$ can be obtained by setting  $\hat{N}_{\phi,\hat{\phi}}=0$, $\forall \phi,\hat{\phi}\in\mathcal{P}$, i.e., $\tilde{\epsilon}=\frac{\tilde{\Theta}\tilde{c}}{1-\mathrm{e}^{\overline{\varsigma}}}$); that \eqref{TDADT_a} and \eqref{TDADT_b} are satisfied for $j=0$ implies that only TDADT switchings need to be performed for the GUAS property. On the other hand, \eqref{V_bounded}, \eqref{V_decay}, \eqref{V_jump} are the Lyapunov-like conditions that serve the similar purposes to those
 in existing works that used MLFs in stability analysis of switched systems (see e.g., (3.8), (3.9), (3.6) of \cite{RN1170}, (2), (3), (5) of \cite{RN1188}, (13)-(15) of \cite{RN1286}). Note that here \eqref{V_jump} further relaxes the existing related conditions (e.g., (15) of \cite{RN1286}) by adding the offset term $\tilde{\Theta}\geq 0$. This allows a more general case where non-vanishing positive jumps exist at some switching instants for $V_{\sigma(t)}$ (which leads to the GUPS of \eqref{MDSS}). Setting $\tilde{\Theta}=0$, then \eqref{V_jump} will reduce to a similar form to, e.g., (15) of \cite{RN1286}.}
\end{remark}

{Despite that Theorem \ref{Thm_1} provides essential stability criteria for the $M^3D$ system \eqref{MDSS}, it has claimed in advance the existence of the function $V_{\phi}$ satisfying the Lyapunov-like conditions \eqref{V_bounded}, \eqref{V_decay}, and \eqref{V_jump} for each $\phi\in\mathcal{P}$ without providing a method to explicitly find such a function. Besides, the conditions such as \eqref{V_decay} and \eqref{V_jump} have not shown sufficient inherent
connections with the subsystem dynamics \eqref{MDSS} of the $M^3D$ system and its state
transition process \eqref{Mode_transform_var}, which leaves these conditions unverified. These have consequently made Theorem \ref{Thm_1} restricted in application. To this end, next we will give some further discussions on the stability of the $M^3D$ system.
Note that in the following the linear subsystem model \eqref{MDSS_linear_closed_loop} will be considered instead of the nonlinear one \eqref{MDSS}, such that more detailed structural information can be exploited in analysis.}

\subsection{Parametric MLFs}\label{section_3_2}
{Following the method for the stability analysis of conventional switched linear systems \cite{RN214}, a potential explicit candidate of $V_{\sigma(t)}$ for the $M^3D$ system \eqref{MDSS_linear_closed_loop} can be constructed into a piecewise quadratic form, i.e., $V_{\sigma(t)}(t,x_{\sigma(t)}(t))=x_{\sigma(t)}^T(t)P_{\sigma(t)}x_{\sigma(t)}(t)$ with a positive definite matrix $P_{\phi}\in\mathbb{R}^{n_{\phi}\times n_{\phi}}$ for each $\phi\in\mathcal{P}$.
Based on it, a series of linear matrix inequalities of $P_{\phi}$ (such as (23) and (24) of \cite{RN1286}) can be established via the Lyapunov-like conditions as \eqref{V_bounded}-\eqref{V_jump}. Then, the existence of the candidate function $V_{\sigma(t)}$ boils down to the existence of $P_{\phi}$, i.e., the feasibility of the associated linear matrix inequalities.
However, such feasibility is usually assumed (instead of verified) to be true in advance, which essentially makes the related stability result for the linear model a special case of that for the nonlinear model (see e.g., \cite[Theorem 2]{zhao2017new}).}
{Given this, and
to further verify the stability conditions in Theorem \ref{Thm_1}, we will introduce to \eqref{MDSS_linear_closed_loop} a new class of MLFs, called the \textit{parametric MLFs}, as an explicit candidate of $V_{\sigma(t)}$ in Theorem \ref{Thm_1}.
}

\subsubsection{Construction of parametric MLFs}
{
Specifically, for any $t\in[t_0,t_f]$, a candidate of {parametric MLFs} for the $M^3D$ system \eqref{MDSS_linear_closed_loop} can be constructed  as
\begin{align}\label{D_MLFs}
V_ {\sigma(t)}(t,x_{\sigma(t)}(t))=\eta(t)x^T_{\sigma(t)}(t)P_{\sigma(t)}x_{\sigma(t)}(t),
\end{align}
 where the time-varying parameter $\eta(t)$ is a bounded right-continuous piecewise constant function of $t$, i.e., $\forall t\in[t_k,t_{k+1})$, $\eta(t)=\eta_k\in[\underline{\eta},\bar{\eta}]$ in which $\eta_k>0$ is constant for each $k\in\mathbb{N}_{\geq 0}$, $\underline{\eta}$ and $\bar{\eta}$ are certain positive constants; for each $\phi\in\mathcal{P}$, $P_{\phi}\in\mathbb{R}^{n_\phi\times n_\phi}$ is a positive definite matrix. Then, it follows from \eqref{Mode_transform_var} that for any $t_k$,
$\|x_{\sigma(t^+_k)}(t^+_k)\|^2\leq  2\|\Xi_{\sigma(t^+_k),\sigma(t^-_k)}\|^2\|x_{\sigma(t^-_k)}(t^-_k)\|^2+2\|\Phi_k\|^2$,
which implies
\begin{align}\label{V2}
V_{\sigma(t_k^+)}(t_k^+,x_{\sigma(t^+_k)}(t^+_k))\leq  & {\Omega}_{\sigma(t^+_k),\sigma(t^-_k)}V_{\sigma(t_k^-)}(t^-_k,x_{\sigma(t^-_k)}(t^-_k))\nonumber\\&+\tilde{\Theta},
\end{align}
in which,
\begin{subnumcases}{}
\label{V_parameters_independent}
{\Omega}_{\phi,\hat{\phi}}=\max\limits_{\forall k\in\Gamma_{\phi,\hat{\phi}}}(\frac{2\eta_k\lambda_{\max}(P_{\phi})}{\eta_{k-1}\lambda_{\min}(P_{\hat{\phi}})}\Upsilon),\ \ \forall \phi,\hat{\phi}\in\mathcal{P},\\ \label{V_parameters_independent111}
\tilde{\Theta}=\tilde{\vartheta},
\end{subnumcases}
where
$\Gamma_{\phi,\hat{\phi}}=\{k|\sigma(t_{k-1})=\hat{\phi},\sigma(t_k)=\phi\}$; $\Upsilon=\max(1,\max\limits_{\forall \phi,\hat{\phi}\in\mathcal{P}}\|{\Xi}_{\phi,\hat{\phi}}\|^2)$,
$\tilde{\vartheta}=2\bar{\eta}\max \limits_{\forall \phi\in\mathcal{P}}\lambda_{\max}(P_{\phi})\|\bar{\Phi}\|^2$
for the state-independent $\Phi_k$; $\Upsilon=\max(1,\max\limits_{\forall \phi,\hat{\phi}\in\mathcal{P}}\|\check{\Xi}_{\phi,\hat{\phi}}\|^2)$, $\tilde{\vartheta}=0$ for the state-dependent $\Phi_k$ that satisfies \eqref{Phi_value}.}
{It can then be seen that \eqref{V2} is consistent in form with \eqref{V_jump} of Theorem \ref{Thm_1}, but the value range of $\Omega_{\phi,\hat{\phi}}$ for each $\phi,\hat{\phi}\in\mathcal{P}$ cannot be ensured from \eqref{V_parameters_independent} without knowing the evolution of $\eta(t)$. Hence, we propose the following procedure for the update of $\eta(t)$ at each $t_k\in[t_0,t_f]$. Note that in the following, ${\tilde{\Delta}}_k\triangleq \frac{2\lambda_{\max}(P_{\sigma(t_k)})}{\lambda_{\min}(P_{\sigma(t_{k-1})})}\Upsilon$, $\forall k\in\{1,...,N(t_0,t_f)\}$}.

{
\begin{procedure}[Update of $\eta(t_k)=\eta_k$, $t_0 \leq t_k \leq t_f$]\label{Proc_1}
${}$\\
{Step 1}: $k\gets0$;
  initialize $\eta_0>0$, $\chi_{\phi,\hat{\phi}}\in(0,1)$, $\forall \phi\in\mathcal{P}$, $\forall \hat{\phi}\in\mathcal{P}$;  \\
{Step 2:} $k\gets k+1$,  $\eta_k\gets\eta_{k-1}$; if $k> N(t_0,t_f)$, then go to Step 6;\\
{Step 3:}  If $\sigma(t_k)\in\mathcal{P}_u$ and ${\tilde{\Delta}}_k\geq1$ (${\tilde{\Delta}}_k<1$), then go to Step 4 (Step 2); if $\sigma(t_k)\in\mathcal{P}_s$ and $\sigma(t_{k-1})\in\mathcal{P}_u$ and $k>1$ and ${\tilde{\Delta}}_{k-1}\geq1$, then go to Step 5, else go to Step 2; \\
{Step 4:}
$\eta_k\gets{\chi_{\sigma(t_k),\sigma(t_{k-1})}\eta_k}/{{\tilde{\Delta}}_k}$, go to Step 2;\\
{Step 5:} $\eta_k\gets\frac{\eta_k\tilde{\Delta}_{k-1}}{\chi_{\sigma(t_{k-1}),\sigma(t_{k-2})}}$, go to Step 2;\\
{Step 6:} Exit.
\end{procedure}}
{Note that in Procedure \ref{Proc_1} we suppose the switching signal $\sigma(t)$ of \eqref{MDSS_linear_closed_loop} to satisfy that for any $t_k$, $\forall k\in\mathbb{N}$, if $\sigma(t_{k-1})\in\mathcal{P}_u$, then $\sigma(t_{k})\in\mathcal{P}_s$. Formally, the switching signal with this property is called \textit{``quasi-alternative''} \cite{RN2201}, and we denote the set of quasi-alternative switching signals by $\tilde{\Psi}_{\sigma}$.}
A quasi-alternative switching signal ensures the destabilizing effect of an unstable subsystem can immediately be compensated by the stabilizing effect of the stable subsystem that follows. Such a compensation method has been commonly employed for dealing with the stability problems of switched systems in the presence of both stable and unstable subsystems, see e.g., \cite{zhao2017new,878825}. {Also note that the employment of the quasi-alternative switching signal indeed brings some restrictions, as it prevents the switching between unstable subsystems which further rules out the case where all the subsystems are unstable \cite{RN1188,RN1159}. We allow such conservativeness for a trade-off purpose considering the potential extra complexity it could bring by studying the stability problem of the $M^3D$ system with all unstable subsystems. For this issue, some more endeavors need to be further made since the compensation method would be less effective in the absence of stable subsystems. This will be one of our main focuses in future works.}

\subsubsection{Behaviors of the time-varying parameter $\eta(t)$}
{The parametric MLFs feature a time-varying parameter $\eta(t)$ subject to Procedure \ref{Proc_1}. It can be readily seen that if $\eta(t)\equiv1$, the parametric MLFs will reduce to the classic quadratic form.
{Note that when $\tilde{\Delta}_{k}\geq 1$ and $0<\chi_{\sigma(t_{k}),\sigma(t_{k-1})}<1$ for some $k\in\mathbb{N}$, $\sigma(t_{k})\in\mathcal{P}_u$, then Step 4 of Procedure \ref{Proc_1} actually indicates a decrease update of $\eta_k$ for the switching from $\hat{\phi}\in\mathcal{P}$ to $\phi\in\mathcal{P}_u$, which implies the condition $0<\Omega_{\phi,\hat{\phi}}<1$ for \eqref{V_jump} can always be satisfied.
However, constant executions of such pure decease updates would potentially drive $\eta_k$ to 0 as $k\rightarrow +\infty$, which makes the parametric MLFs trivial to use. To avoid this, the switching signal here for \eqref{MDSS_linear_closed_loop} is thus required to satisfy $\sigma(t)\in\tilde{\Psi}_{\sigma}$, i.e., an unstable subsystem must be followed by a stable one. In this case, once a decrease update of $\eta(t)$ (Step 4) is made at a switching instant, an increase update (Step 5) which is identical in magnitude to the decrease update will be made immediately at the next switching instant. Such an update procedure can thus confine the value of $\eta(t)$ to a bounded range of $[\underline{\eta},\overline{\eta}]$, where
$\underline{\eta}={\eta_0}\min\limits_ {\forall\phi,\hat{\phi}\in\mathcal{P}}\chi_{\phi,\hat{\phi}}/(2\Upsilon{\max\limits_ {\forall\phi,\hat{\phi}\in\mathcal{P}}{\lambda_{\max}(P_{\phi})}/{\lambda_{\min}(P_{\hat{\phi}})}})$, $\bar{\eta}=\eta_0$. Note that here ${\max\limits_ {\forall\phi,\hat{\phi}\in\mathcal{P}}{\lambda_{\max}(P_{\phi})}/{\lambda_{\min}(P_{\hat{\phi}})}}>1$ and ${{\Upsilon}}\geq 1$}. Moreover, for each $\phi\in\mathcal{P}_u$, $\hat{\phi}\in\mathcal{P}_s$, $\chi_{\phi,\hat{\phi}}$ is a given positive scalar that affects the range of $\eta(t)$.}

With the proposed parametric MLFs, we are going to present the stability result for the linear $M^3D$ system \eqref{MDSS_linear_closed_loop}.
\subsection{Stability of linear $M^3D$ system via parametric MLFs}\label{section_3_3}
 The following theorem provides the stability conditions for \eqref{MDSS_linear_closed_loop} based on the proposed parametric MLFs.
\begin{thm}\label{thm_linear_MDSS}
{Consider the switched system \eqref{MDSS_linear_closed_loop} with the state transition \eqref{Mode_transform_var} at $t_k$, $k\in\mathbb{N}$. Given
$\eta_0>0$, $\chi_{\phi,\hat{\phi}}\in(0,1)$, $\phi\in\mathcal{P}_u$, $\hat{\phi}\in\mathcal{P}$, $\sigma(t)\in\tilde{\Psi}_{\sigma}$, and state-independent $\Phi_k$, if conditions \eqref{TDADT_a} and \eqref{TDADT_b} hold, where $\Omega_{\phi,\hat{\phi}}$ satisfies \eqref{V_parameters_independent} under Procedure \ref{Proc_1}, and $\tilde{\gamma}_{\phi}=2\gamma_{\phi}$ with
 \begin{subnumcases}{\label{tilde_gamma_111}}
 \max\limits_{\forall j}{\mathrm{Re}}(\lambda_j(\tilde{A}_{\phi}))<\gamma_{\phi}<0, \ \forall \phi\in\mathcal{P}_s,  \label{tilde_gamma}\\
{
 0 \leq \max\limits_{\forall  j}{\mathrm{Re}}(\lambda_j(\tilde{A}_{\phi}))<\gamma_{\phi}, \
  \forall \phi\in\mathcal{P}_u,} \label{tilde_gamma1}
  \end{subnumcases}
in which $\tilde{A}_{\phi}=A_{\phi}+B_{\phi}K_{\phi}$, $\forall \phi\in\mathcal{P}$,
then \eqref{MDSS_linear_closed_loop} is GUPS with $\epsilon=\sqrt{\tilde{\epsilon}\underline{\eta}^{-1}\max\limits_{\forall \phi\in\mathcal{P}}(1/\lambda_{\min}(P_{\phi}))}$, where $\tilde{\epsilon}$ is given as in Theorem \ref{Thm_1} with $\tilde{\Theta}$ satisfying \eqref{V_parameters_independent111}, $\underline{\eta}={\eta_0}\min\limits_ {\forall \phi,\hat{\phi}\in\mathcal{P}}\chi_{\phi,\hat{\phi}}/(2\Upsilon{\max\limits_ {\forall \phi,\hat{\phi}\in\mathcal{P}}{\lambda_{\max}(P_{\phi})}/{\lambda_{\min}(P_{\hat{\phi}})}})$, and $P_{\phi}$, $\forall \phi\in\mathcal{P}$ satisfies
\begin{align}\label{Lyapunov_eqn}
\tilde{A}_{\phi}^TP_{\phi}+P_{\phi}\tilde{A}_{\phi}-2\gamma_{\phi}P_{\phi}=-I_{n_{\phi}}.
\end{align}
Particularly, if \eqref{TDADT_a} and \eqref{TDADT_b} are satisfied for $j=0$ and state-dependent $\Phi_k$,  then \eqref{MDSS_linear_closed_loop} is GUAS.}
 \end{thm}
\begin{proof}\upshape
For $\sigma(t)\in\tilde{\Psi}_{\sigma}$ and any $t\in[t_{k},t_{k+1})$, $k=0,1,...,N(t_0,t_f)$, construct a parametric MLFs candidate for \eqref{MDSS_linear_closed_loop} as in \eqref{D_MLFs}, where $P_{\phi}$ satisfies the equation \eqref{Lyapunov_eqn} for each $\phi\in\mathcal{P}$. Clearly, for $\gamma_{\phi}$, $\phi\in\mathcal{P}$ satisfying \eqref{tilde_gamma} and \eqref{tilde_gamma1}, \eqref{Lyapunov_eqn} becomes a Lyapunov equation which implies that $P_{\phi}$ is a positive definite solution for each $\phi\in\mathcal{P}$. It can thus be derived from \eqref{D_MLFs} and Procedure \ref{Proc_1} that $\underline{\eta}\min\limits_{\forall \phi\in\mathcal{P}}\lambda_{\min}(P_{\phi})\|x_{\sigma(t)}(t)\|^2\leq V_ {\sigma(t)}(t,x_{\sigma(t)}(t))\leq \overline{\eta}\max\limits_{\forall \phi\in\mathcal{P}}\lambda_{\max}(P_{\phi})\|x_{\sigma(t)}(t)\|^2$,
which means \eqref{V_bounded} is verified. Then, by \eqref{D_MLFs},  \eqref{tilde_gamma}, \eqref{tilde_gamma1}, \eqref{Lyapunov_eqn}, one gets  for any $\phi\in\mathcal{P}$,
\begin{gather}\label{Lyapunov_1}
\dot{V}_{\phi}(t,x_{\phi}(t))\leq\tilde{\gamma}_{\phi}V_{\phi}(t,x_{\phi}(t)),
\end{gather}
where $\tilde{\gamma}_{\phi}=2\gamma_{\phi}$, which verifies \eqref{V_decay}. Moreover, by \eqref{Mode_transform_var} and \eqref{D_MLFs}, one has that
 \eqref{V2} holds at any switching instant $t_k$ with ${\Omega}_{\phi,\hat{\phi}}$ and $\tilde{\Theta}$ satisfying \eqref{V_parameters_independent} and \eqref{V_parameters_independent111}, respectively, which implies that for $\phi=\sigma(t_k^+)\in\mathcal{P}$, $\hat{\phi}=\sigma(t_{k}^-)\in\mathcal{P}$:
\begin{align}\label{Mode_transform_V}
V_{\phi}(t_k^+,x_{\phi}(t_k^+))\leq  {\Omega}_{\phi,\hat{\phi}}V_{\hat{\phi}}(t_k^-,x_{\hat{\phi}}(t_k^-))+\tilde{\Theta}.
\end{align}
Here, note that the update of $\eta_k$ by Procedure \ref{Proc_1} always guarantees the condition $0<\Omega_{\phi,\hat{\phi}}<1$ for any $\phi\in\mathcal{P}_u$, $\hat{\phi}\in\mathcal{P}_s$. Meanwhile, \eqref{V_parameters_independent} and Procedure \ref{Proc_1} indicate that: when $\phi\in\mathcal{P}_s$ and $\hat{\phi}\in\mathcal{P}_u$, there holds $\Omega_{\phi,\hat{\phi}}\leq\bar{\Omega}_{\phi,\hat{\phi}}=\frac{4\lambda_{\max}(P_{\phi})\lambda_{\max}(P_{\hat{\phi}})}{\underline{\chi}\min\limits_{\forall\phi\in\mathcal{P}}(\lambda_{\min}(P_{\phi}))\lambda_{\min}(P_{\hat{\phi}})}\Upsilon^2$ with $\underline{\chi}=\min\limits_{\forall\phi,\hat{\phi}\in\mathcal{P}}\chi_{\phi,\hat{\phi}}$, which implies $\bar{\Omega}_{\phi,\hat{\phi}}>1$, for any $\phi\in\mathcal{P}_s$, $\hat{\phi}\in\mathcal{P}_u$; when $\phi,\hat{\phi}\in\mathcal{P}_s$, there holds $\Omega_{\phi,\hat{\phi}}\leq \bar{\Omega}_{\phi,\hat{\phi}}=\tilde{F}(\frac{2\lambda_{\max}(P_{\phi})}{\lambda_{\min}(P_{\hat{\phi}})}\Upsilon,\max\limits_{\forall\phi,\hat{\phi}\in\mathcal{P}}\frac{2\lambda_{\max}(P_{\phi})}{\lambda_{\min}(P_{\hat{\phi}})}\Upsilon)$, $\tilde{F}(a,b):\mathbb{R}\times\mathbb{R}\rightarrow\mathbb{R}$ satisfies that if $a>1$, then $\tilde{F}(a,b)=a$, else $\tilde{F}(a,b)=b$, which ensures $\bar{\Omega}_{\phi,\hat{\phi}}>1$ for any $\phi\in\mathcal{P}_s$, $\hat{\phi}\in\mathcal{P}_s$. This means the value of $\Omega_{\phi,\hat{\phi}}$ in \eqref{Mode_transform_V} whenever  $\phi\in\mathcal{P}_s$ and $\hat{\phi}\in\mathcal{P}$ can be replaced by $\bar{\Omega}_{\phi,\hat{\phi}}$, which guarantees $\Omega_ {\phi,\hat{\phi}}>1$ for any $\phi\in\mathcal{P}_s$ and $\hat{\phi}\in\mathcal{P}$, and further verifies \eqref{V_jump}. Then, for state-independent $\Phi_k$, one concludes from Theorem \ref{Thm_1} that with \eqref{TDADT_a} and \eqref{TDADT_b} satisfied, there holds $\limsup\limits_{t\rightarrow +\infty}V_ {\sigma(t)}(t)\leq\tilde{\epsilon}$, with $\tilde{\epsilon}$ given as in Theorem \ref{Thm_1}. Moreover, by Procedure \ref{Proc_1} one has that  $0<\eta_k^{-1}\leq \underline{\eta}^{-1}$, which along with \eqref{V_bounded} and \eqref{D_MLFs} further denotes that $\limsup\limits_{t\rightarrow +\infty}\|x_{\sigma(t)}(t)\|\leq \epsilon$, $\epsilon=\sqrt{\tilde{\epsilon}\underline{\eta}^{-1}\max\limits_{\forall \phi\in\mathcal{P}}(1/\lambda_{\min}(P_{\phi}))}$.
This implies \eqref{MDSS_linear_closed_loop} is GUPS. Particularly, for state-dependent $\Phi_k$ satisfying \eqref{Phi_value}, it follows from \eqref{V_parameters_independent111} that $\tilde{\Theta}=0$, then with \eqref{TDADT_a} and \eqref{TDADT_b} satisfied for $j=0$, one concludes from Theorem \ref{Thm_1} that $\lim\limits_{t\rightarrow +\infty}V_{\sigma(t)}(t)=0$. This along with \eqref{V_bounded} and \eqref{D_MLFs} indicates that $\lim\limits_{t\rightarrow +\infty}\|x_{\sigma(t)}(t)\|=0$, i.e., \eqref{MDSS_linear_closed_loop} is GUAS.$\blacksquare$
\end{proof}

\begin{remark}\upshape
Compared with Theorem \ref{Thm_1}, Theorem \ref{thm_linear_MDSS} further reveals the connections between the two stability properties of the $M^3D$ system and the two types of the impulse $\Phi_k$ introduced in Section \ref{section_2_3}. Specifically, for the state-independent impulse $\Phi_k$, due to its non-vanishing property that potentially impedes an asymptotic convergence of the state, the GUPS of \eqref{MDSS_linear_closed_loop} is sought in Theorem \ref{thm_linear_MDSS}, which corresponds to the general case of $\tilde{\Theta}\geq 0$ in Theorem \ref{Thm_1}. For state-dependent $\Phi_k$, owing to its vanishing property, the GUAS of \eqref{MDSS_linear_closed_loop} can be ensured, which corresponding to the special case of $\tilde{\Theta}=0$ in Theorem \ref{Thm_1}. {Moreover, Theorem \ref{thm_linear_MDSS} also verifies the Lyapunov-like conditions \eqref{V_bounded}, \eqref{V_decay}, \eqref{V_jump} in Theorem \ref{Thm_1}. As is shown in the proof of Theorem \ref{thm_linear_MDSS}, all these Lyapunov-like conditions are derived from the subsystem dynamics of \eqref{MDSS_linear_closed_loop} and the state transition process \eqref{Mode_transform_var} under the constructed parametric MLFs \eqref{D_MLFs}, which means they are naturally satisfied for the linear $M^3D$ system \eqref{MDSS_linear_closed_loop}.} {In this case, one only needs to ensure for \eqref{MDSS_linear_closed_loop} the TDADT conditions \eqref{TDADT_a} and \eqref{TDADT_b} of Theorem \ref{Thm_1} which reflect typical time-dependent switching methods that are usually easy to realize in practice \cite{zhao2017new,RN637}.}
\end{remark}

{It is notable that the stability of the classic switched systems has been widely used in applications such as the cooperative control of MASs with switching features \cite{RN648,RN1281,RN921}. In these applications, the time-dependent switching methods have been commonly adopted (e.g., the dwell-time method used in \cite{RN1281}, \cite{RN921}; the ADT method used in \cite{RN648}). Given these facts, it is then of interest to seek a potential application of the results obtained for the $M^3D$ system.}

\section{Application to consensus of open MASs}\label{section_4}
{In this section, we are going to indicate a potential application of the stability results for $M^3D$ systems to consensus of open MASs. Open MASs can cover a wide range of emerging real-world networked systems with a varying size and node number, such as the vehicle platoons with lane change maneuvers \cite{852914}, and the social networks \cite{RN2223}.}
Note that for brevity and consistency, some notations for switching properties (e.g., $\sigma(t)$, $\mathcal{P}$) in the previous sections will be reused for the open MAS.
\subsection{System formulation and preliminaries}\label{section_4_1}
 The considered interaction topology is characterized by a digraph $\mathcal{G}_{\sigma(t)}=\{\mathcal{V}_{\sigma(t)},\mathcal{E}_{\sigma(t)}\}$ under the switching signal $\sigma(t)$, $\sigma:\mathbb{R}_{\geq0}\rightarrow \mathcal{P}=\{1,2,...,s\}$, where $s$ is a finite positive integer, $\mathcal{V}_{\sigma(t)}=\{1,2,...,N_{\sigma(t)}\}$ denotes the {label set of vertices} of $\mathcal{G}_{\sigma(t)}$, $\mathcal{E}_{\sigma(t)}\subseteq\mathcal{V}_{\sigma(t)}\times \mathcal{V}_{\sigma(t)}$ denotes the edge set of $\mathcal{G}_{\sigma(t)}$.
Denote by $\mathcal{A}_{\sigma(t)}=[a_{ij}(\sigma(t))]\in\mathbb{R}^{N_{\sigma(t)}\times N_{\sigma(t)}}$, $\forall i,j\in\mathcal{V}_{\sigma(t)}$ the adjacency matrix of $\mathcal{G}_{\sigma(t)}$, in which $a_{ij}(\sigma(t))=0$ if $(j,i)\not\in\mathcal{E}_{\sigma(t)}$, i.e., there is no directed edge from $j$ to $i$, otherwise $a_{ij}(\sigma(t))=1$, and suppose that $a_{ii}(\sigma(t))=0$ for any $t$ and $i\in\mathcal{V}_{\sigma(t)}$, i.e., no self-loops. The corresponding Laplacian matrix of $\mathcal{G}_{\sigma(t)}$ is denoted by $L_{\sigma(t)}=[l_{ij}(\sigma(t))]\in\mathbb{R}^{N_{\sigma(t)}\times N_{\sigma(t)}}$, $l_{ij}(\sigma(t))=-a_{ij}(\sigma(t))$, $l_{ii}(\sigma(t))=\sum\limits_{j=1}^{N_{\sigma(t)}}a_{ij}(\sigma(t))$, $\forall i\neq j$, $i,j\in\mathcal{V}_{\sigma(t)}$.

Under the above topology setting, consider the open MAS with linear agent dynamics and a distributed linear consensus controller for each $i\in\mathcal{V}_{\sigma(t)}$:
\begin{align}\label{open_MAS}
\dot{\xi}_{\sigma(t),i}(t)=&S_{}\xi_{\sigma(t),i}(t)-\varrho\sum\limits_{j=1}^{N_{\sigma(t)}}a_{ij}{(\sigma(t))}(\xi_{\sigma(t),i}(t)\nonumber\\&-\xi_{\sigma(t),j}(t)),\ \end{align}
where $\varrho>0$ is a given scalar; the state of the agent labeled $i$ is denoted by ${\xi}_{\sigma(t),i}(t)=[{\xi}^1_{\sigma(t),i}(t),...,{\xi}^p_{\sigma(t),i}(t)]^T\in\mathbb{R}^p$, where ${\xi}^j_{\sigma(t),i}(t)\in\mathbb{R}$ is the $j$-th component of ${\xi}_{\sigma(t),i}(t)$; $S\in\mathbb{R}^{p\times p}$ with $\min\limits_{\forall j}\mathrm{Re}(\lambda_j(S))\geq 0$. A compact form of \eqref{open_MAS} can be given by:
\begin{align}\label{open_MAS_comp}
\dot{\tilde{\xi}}_{\sigma(t)}(t)=(I_{N_{\sigma(t)}}\otimes S-\varrho L_{\sigma(t)}\otimes I_p)\tilde{\xi}_{\sigma(t)}(t),
 \end{align}
where $\tilde{\xi}_{\sigma(t)}(t)=[\xi^T_{\sigma(t),1}(t),...,\xi^T_{\sigma(t),N_{\sigma(t)}}(t)]^T\in\mathbb{R}^{pN_{\sigma(t)}}$.

The similar agent dynamics to \eqref{open_MAS} can also be found in, e.g., \cite{RN2273}.
{Note that although there is no concrete practical background specified for the considered open MAS, the linear models like \eqref{open_MAS}, \eqref{open_MAS_comp} can still effectively approximate various practical MASs, such as the cooperative unmanned vehicles \cite{RN3375,RN648,RN1281,7275192}. Besides, the linear consensus controller as in \eqref{open_MAS} is also considered convenient for hardware implementations in practice \cite{RN3375}.}

The agent migration behavior of the considered open MAS \eqref{open_MAS_comp} at each switching instant $t_k$ of $\sigma(t)$ is captured as the following state transition process of $\tilde{\xi}_{\sigma(t)}(t)$:
\begin{align}\label{Agent_moves}
\tilde{\xi}_{\sigma(t_k^+)}(t_{k}^+)=&\bar{\tilde{\Xi}}_{\sigma(t_k^+),\sigma(t_k^-)}\tilde{\xi}_{\sigma(t^-_k)}(t_{k}^-)+\bar{\tilde{\Phi}}_k,
\end{align}
where $\bar{\tilde{\Xi}}_{\sigma(t_k^+),\sigma(t_k^-)}\in\mathbb{B}^{pN_{\sigma(t_k^+)}\times pN_{\sigma(t_k^-)}}$ and $\bar{\tilde{\Phi}}_k\in\mathbb{R}^{pN_{\sigma(t_k^+)}}$. In particular, we have the following summary of specific agent migration behaviors reflected by \eqref{Agent_moves}, which is in most part consistent with \cite{hendrickx2017open} and links with Section \ref{section_2_2}:
\begin{enumerate}
  \item \textbf{Arrival}: There are new agents joining the original group, i.e., $|\mathcal{V} _{\sigma(t_k^-)}|<|\mathcal{V}_{\sigma(t_k^+)}|$, $N_{\sigma(t_k^-)}<N_{\sigma(t_k^+)}$. The joined agents can instantly establish new connections with other existing agents. For a pure arrival behavior of an agent, the matrix $\bar{\tilde{\Xi}}_{\sigma(t_k^+),\sigma(t_k^-)}$ is obtained by inserting a $p\times pN_{\sigma(t_k^-)}$ zero matrix between specific rows, say, the $h$-th row and the $h+1$-th row of $I_{pN_{\sigma(t_k^-)}}$, where $h$ is a given non-negative multiple of $p$; correspondingly, the vector $\bar{\tilde{\Phi}}_k$ is obtained by inserting a $p\times 1$ vector denoting the state values of the joined agent between the $h$-th and the $h+1$-th entries of a $pN_{\sigma(t_k^-)}\times 1$ zero vector.
  \item \textbf{Departure}: There are agents leaving the original group, i.e., $|\mathcal{V}_{\sigma(t_k^-)}|>|\mathcal{V}_{\sigma(t_k^+)}|$, $N_{\sigma(t_k^-)}>N_{\sigma(t_k^+)}$. Once an agent has left the group, any connections originally associated with it will lose. For a pure departure behavior of an agent, $\bar{\tilde{\Xi}}_{\sigma(t_k^+),\sigma(t_k^-)}$ is derived by removing a $p\times pN_{\sigma(t_k^-)}$ matrix from the specific position of $I_{pN_{\sigma(t_k^-)}}$. Clearly, there holds $\bar{\tilde{\Phi}}_k\equiv 0$ for this case if no self impulse of the agent state exists.
  \item \textbf{Replacement}: There are agents instantly replaced by new agents. A replacement behavior can also be deemed a simultaneous occurrence of the arrival and departure behaviors. A pure replacement behavior does not change the size of the network topology, i.e., $|\mathcal{V}_{\sigma(t_k^+)}|=|\mathcal{V}_{\sigma(t_k^-)}|$, $N_{\sigma(t_k^-)}=N_{\sigma(t_k^+)}$, which implies $\bar{\tilde{\Xi}}_{\sigma(t_k^+),\sigma(t_k^-)}\equiv I_{pN_{\sigma(t_k^-)}}$. The instantaneous state variation of the replaced agent is brought solely by $\bar{\tilde{\Phi}}_k$.
\end{enumerate}

Since the vector $\bar{\tilde{\Phi}}_k$ of \eqref{Agent_moves} also indicates the impulsive effect of the agent state, one can thus specify its state-independent or state-dependent property as in Section \ref{section_2_3}. Particularly, similar to \eqref{Phi_value},  define state-dependent $\bar{\tilde{\Phi}}_k$ as $\bar{\tilde{\Phi}}_k=\bar{\hat{\Xi}}_{\sigma(t_k^+),\sigma(t_k^-)}\tilde{\xi}_{\sigma(t_k^-)}(t_k^-)$, where $\bar{\hat{\Xi}}_{\sigma(t_k^+),\sigma(t_k^-)}=\tilde{\hat{\Xi}}_{\sigma(t_k^+),\sigma(t_k^-)}\tilde{\Upsilon}_{\sigma(t_k^-)}-\bar{\tilde{\Xi}}_{\sigma(t_k^+),\sigma(t_k^-)}+\widehat{\Upsilon}_{\sigma(t_k^+),\sigma(t_k^-)}$, in which $\tilde{\hat{\Xi}}_{\sigma(t_k^+),\sigma(t_k^-)}\in\mathbb{R}^{pN_{\sigma(t_k^+)}\times p(N_{\sigma(t_k^-)}-1)}$ is a given matrix, $\tilde{\Upsilon}_{\sigma(t)}=[I_{N_{\sigma(t)}-1},-\textbf{1}_{N_{\sigma(t)}-1}]\otimes I_p$, $\widehat{\Upsilon}_{\sigma(t^+),\sigma(t^-)}=[\textbf{0}_{N_{\sigma(t^+)}\times (N_{\sigma(t^-)}-1)},\textbf{1}_{N_{\sigma(t^+)}}]\otimes I_p$ with $\textbf{0}_{m\times n}$ denoting an $m\times n$ matrix of zeros.

\begin{remark}\upshape
{The agent migration behaviors depicted above indicate that the interaction topology of the open MAS \eqref{open_MAS_comp}, represented by $\mathcal{G}_{\sigma(t)}$, is inherently switching and size-varying.} {Meanwhile, they can also well reflect some practical situations, such as the lane change maneuvers in vehicle platoons, which cause the vehicles to join/leave a platoon \cite{852914}.} {Note that for the vertex (label) set $\mathcal{V}_{\sigma(t)}$, we always keep a continuous labeling of agents starting from 1, while also allow the same label to indicate different agents at different times such that no loss of generality will be caused. Also note that it is possible to use a fixed-size switching graph to describe an open MAS network, given that its maximum capacity, i.e., the total number of nodes allowed by the network, is fixed and known.
For this case, the agent who leaves/joins the group can be regarded as a fixed node that loses/regains connections with others. However, in practice it is usually unnecessary to determine an upper bound of the node number, especially when the network scale is unpredictably increasing. Besides, when the singleton nodes largely outnumber the connected nodes, it would be less efficient to consider the dynamics of all these singleton nodes in computation since they contribute little to the evolution of the whole open MAS. This thus necessitates the use of the size-varying graph to describe the open MAS network.}
\end{remark}

For the considered open MAS \eqref{open_MAS_comp}, we are interested in the following two types of consensus performances:
\begin{definition}\label{consensus_GUPS}
{For the open MAS \eqref{open_MAS_comp}, it is said to achieve practical consensus if there exists
$\varepsilon\geq0$ such that
\begin{equation}\label{consensus_p}
\limsup\limits_{t\rightarrow +\infty}\|\xi_{\sigma(t),i}(t)-\xi_{\sigma(t),j}(t)\|\leq\varepsilon, ~~\forall i,j\in \mathcal{V}_{\sigma(t)}.
\end{equation}
Particularly, if \eqref{consensus_p} holds for $\varepsilon=0$, then \eqref{open_MAS_comp} is said to achieve (asymptotic) consensus.}
\end{definition}

Similar definitions for the practical consensus can also be found in \cite{RN1239,7562413}.

The following lemmas are for the upcoming analysis.

\begin{lemma}[\cite{RN648}]\label{MAS_tra}
{For any $\xi_{1},...,\xi_{N} \in\mathbb{R}^p$ and the Laplacian matrix $L=[l_{ij}]$ of a digraph $\mathcal{G}$, there holds:
$\sum\limits_{j=1}^{N}l_{ij}\xi_{j}-\sum\limits_{j=1}^{N}l_{Nj}\xi_{j}=\sum\limits_{j=1}^{N-1}z_{ij}(\xi_{j}-\xi_{N})$,
where $z_{ij}=l_{ij}-l_{Nj}$ for $i,j=1,...,N-1$. Denoting $Z=[z_{ij}]$, then the real parts of all the eigenvalues of $Z$ are non-negative. Moreover, their real parts are all positive provided that $\mathcal{G}$ contains a directed spanning tree.}
\end{lemma}

Note that hereunder we will slightly abuse the notations by letting $\mathcal{P}_s=\{\phi|\mathcal{G}_{\phi} \mbox{ contains a directed spanning tree}\}$, $\mathcal{P}_u=\{\phi|\mathcal{G}_{\phi} \mbox{ contains no directed spanning tree}\}$.

\begin{lemma}[\cite{1084534}]\label{MAS_tra1}
For given matrices $A\in\mathbb{C}^{n\times n}$ and $B\in\mathbb{C}^{r\times r}$, if $Y=A\otimes I_r+I_n\otimes B$, then $\lambda(Y)=\{\lambda_A+\lambda_B|\lambda_A\in\lambda(A),\lambda_B\in\lambda(B)\}$.
\end{lemma}

Detailed proofs for the above two lemmas can  be found in \cite{RN648} (and the references therein) and \cite{bernstein2005matrix,1084534}, respectively.

\subsection{An $M^3D$ system interpretation of open MAS}\label{section_4_2}
In this subsection, we will provide an interpretation of the open MAS \eqref{open_MAS_comp} based on the $M^3D$ system \eqref{MDSS_linear_closed_loop}.

Denote $z_{ij}(\sigma(t))=l_{ij}(\sigma(t))-l_{N_{\sigma(t)}j}(\sigma(t))$ for $i,j=1,...,N_{\sigma(t)}-1$ and $Z_{\sigma(t)}=[z_{ij}(\sigma(t))]$. Then, one can transform \eqref{open_MAS} into the following consensus error system by defining $\delta_{\sigma(t),i}(t)\triangleq\xi_{\sigma(t),i}(t)-\xi_{\sigma(t),N_{\sigma(t)}}(t)$
for $i=1,...,N_{\sigma(t)}-1$:
\begin{align}\label{open_MAS_delta}
\dot{\delta}_{\sigma(t),i}(t)=S\delta_{\sigma(t),i}(t)- \varrho \sum\limits_{j=1}^{N_{\sigma(t)}-1}z_{ij}(\sigma(t))\delta_{\sigma(t),j}(t).
\end{align}
 Further, \eqref{open_MAS_delta} can be rewritten in a compact form:
\begin{align}\label{open_MAS_delta_compact}
\dot{\tilde{\delta}}_{\sigma(t)}(t)=(I_{N_{\sigma(t)}-1}\otimes S-\varrho Z_{\sigma(t)}\otimes I_p)\tilde{\delta}_{\sigma(t)}(t),
\end{align}
where $\tilde{\delta}_{\sigma(t)}(t)=[\delta^T_{\sigma(t),1}(t),...,\delta^T_{\sigma(t),N_{\sigma(t)}-1}(t)]^T\in\mathbb{R}^{p(N_{\sigma(t)}-1)}$ is the error state. It then follows from \eqref{Agent_moves} that
\begin{align}\label{Agent_disagreement_moves}
\tilde{\delta}_{\sigma(t_k^+)}(t_{k}^+)=&\tilde{\Xi}_{\sigma(t_k^+),\sigma(t_k^-)}\tilde{\delta}_{\sigma(t^-_k)}(t_{k}^-)+\tilde{\Phi}_k,
\end{align}
where, $\tilde{\Xi}_{\sigma(t_k^+),\sigma(t_k^-)}\in\mathbb{B}^{{p(N_{\sigma(t^+_k)}-1)\times p(N_{\sigma(t^-_k)}-1)}}$ is obtained by removing the last $p$ rows and the last $p$ columns from $\bar{\tilde{\Xi}}_{\sigma(t_k^+),\sigma(t_k^-)}$; $\tilde{\Phi}_k=\check{\Phi}_k+\hat{\Phi}_k$ with $\tilde{\Phi}_k,\check{\Phi}_k,\hat{\Phi}_k\in\mathbb{R}^{p(N_{\sigma(t^+_k)}-1)}$,
$\check{\Phi}_k=\tilde{\Upsilon}_{\sigma(t_k^+)}\bar{\tilde{\Xi}}_{\sigma(t_k^+),\sigma(t_k^-)}\tilde{\xi}_{\sigma(t_k^-)}(t_k^-)-\tilde{\Xi}_{\sigma(t_k^+),\sigma(t_k^-)}\tilde{\delta}_{\sigma(t_k^-)}(t_k^-)$, $\hat{\Phi}_k=\tilde{\Upsilon}_{{\sigma(t^+_k)}}\bar{\tilde{\Phi}}_k$. Note that
$\tilde{\Phi}_k$ carries two types of impulses: the impulse ${\check{\Phi}}_k$ whose value depends on those of both $\tilde{\xi}_{\sigma(t_k^-)}(t_k^-)$ and $\tilde{\delta}_{\sigma(t_k^-)}(t_k^-)$; the impulse $\hat{\Phi}_k$ whose state dependency is consistent with that of $\bar{\tilde{\Phi}}_k$.
In addition, assume that for any state-independent $\bar{\tilde{\Phi}}_k$ and any $k$, $\|\tilde{\Phi}_k\|\leq \bar{\Phi}$.
For state-dependent $\bar{\tilde{\Phi}}_k$, we have $\tilde{\Phi}_k=(\tilde{\Upsilon}_{{\sigma(t^+_k)}}\tilde{\hat{\Xi}}_{\sigma(t_k^+),\sigma(t_k^-)}-\tilde{\Xi}_{\sigma(t_k^+),\sigma(t_k^-)})\tilde{\delta}_{\sigma(t_k^-)}(t_k^-)$, which means $\tilde{\Phi}_k$ is also state-dependent, and \eqref{Agent_disagreement_moves} becomes
\begin{align}\label{Agent_disagreement_moves_asymp}
\tilde{\delta}_{\sigma(t_k^+)}(t_{k}^+)=\check{\tilde{\Xi}}_{\sigma(t_k^+),\sigma(t_k^-)}\tilde{\delta}_{\sigma(t^-_k)}(t_{k}^-),
\end{align}
where $\check{\tilde{\Xi}} _{\sigma(t_k^+),\sigma(t_k^-)}=\tilde{\Upsilon}_{{\sigma(t^+_k)}}\tilde{\hat{\Xi}}_{\sigma(t_k^+),\sigma(t_k^-)}$. Note that \eqref{Agent_disagreement_moves_asymp} is consistent in form with \eqref{Mode_transform_var_asymp}.
Then, comparing \eqref{open_MAS_delta_compact} and \eqref{Agent_disagreement_moves} with \eqref{MDSS} and \eqref{Mode_transform_var}, one readily concludes that \textit{the consensus error system \eqref{open_MAS_delta_compact} is an $M^3D$ system} with the state transition at the switching instant $t_k$ given by \eqref{Agent_disagreement_moves}.

Next, we will resort to both Lemma \ref{MAS_tra} and Lemma \ref{MAS_tra1} to explore the relations between the connectivity of the digraph $\mathcal{G}_{\phi}$ of the open MAS \eqref{open_MAS_comp} and the stability of the subsystem $\phi$ of the $M^3D$ system \eqref{open_MAS_delta_compact}. Such relations are summarized by the following proposition.

\begin{proposition}\label{MAS_tra2}
    {
    For given matrices $S$, $\min\limits_{\forall i}\mathrm{Re}(\lambda_i(S))\geq 0$, and $Z_{\phi}$, $\phi\in\mathcal{P}$ derived by Lemma \ref{MAS_tra}, if $\phi\in\mathcal{P}_s$, then there exists a constant $\varrho>0$, such that the matrix $I_{N_{\phi}-1}\otimes S-\varrho Z_{\phi}\otimes I_p$ of \eqref{open_MAS_delta_compact} is Hurwitz. Under the same constant $\varrho$, if $\phi\in\mathcal{P}_u$, then $I_{N_{\phi}-1}\otimes S-\varrho Z_{\phi}\otimes I_p$ is non-Hurwitz.}
\end{proposition}
\noindent\textbf{Proof of Proposition \ref{MAS_tra2} }{For ${\phi}\in \mathcal{P}_s$, one concludes from Lemma \ref{MAS_tra} that all the eigenvalues of $Z_{\phi}$ have positive real parts. Recalling that $\min\limits_{\forall i}\mathrm{Re}(\lambda_i(S))\geq 0$ and applying Lemma \ref{MAS_tra1}, it is straightforward to derive that $\max\limits_{\forall i}\mathrm{Re}(\lambda_i(I_{N_{\phi}-1}\otimes S-\varrho Z_{\phi}\otimes I_p))=\max\limits_{\forall i,j}\mathrm{Re}(\lambda_i(S)-\varrho\lambda_j(Z_{\phi}))$. Obviously, one can always find a large enough $\varrho>0$, such that $\max\limits_{\forall i,j}\mathrm{Re}(\lambda_i(S)-\varrho\lambda_j(Z_{\phi}))<0$, i.e., $I_{N_{\phi}-1}\otimes S-\varrho Z_{\phi}\otimes I_p$ is Hurwitz. Moreover, under the same constant $\varrho$, for $\phi\in\mathcal{P}_u$ one can derive from Lemma \ref{MAS_tra} that at least one eigenvalue of $Z_{\phi}$ has a zero real part. By Lemma \ref{MAS_tra1} and $\min\limits_{\forall i}\mathrm{Re}(\lambda_i(S))\geq 0$, it further implies that at least one eigenvalue of $I_{N_{\phi}-1}\otimes S-\varrho Z_{\phi}\otimes I_p$ has a non-negative real part, i.e., the matrix is non-Hurwitz.$\blacksquare$}

\begin{remark}\upshape
Proposition \ref{MAS_tra2} reveals that under a proper $\varrho>0$, the connected/disconnected property of the topology $\mathcal{G}_{\phi}$ of the open MAS \eqref{open_MAS_comp} corresponds to the stable/unstable property of the subsystem $\phi$ of the $M^3D$ system \eqref{open_MAS_delta_compact}. Such a correspondence enables one to seek a further relation between the consensus of open MASs with disconnected digraphs and the stability of $M^3D$ systems with unstable subsystems. This leads to the result that follows.
\end{remark}
\subsection{Consensus of the open MAS via $M^3D$ system stability}\label{section_4_3}
The following theorem summarizes the conditions for the open MAS \eqref{open_MAS_comp} to reach the consensus performances in Definition \ref{consensus_GUPS} via the stability results for the $M^3D$ system.
{
\begin{thm}
\label{exo_consensus}
{The open MAS \eqref{open_MAS_comp} with a size-varying switching digraph $\mathcal{G}_{\sigma(t)}$ can reach practical consensus, if the consensus error system \eqref{open_MAS_delta_compact} is GUPS. The corresponding ultimate bound is given by $\varepsilon=2\sqrt{\tilde{\epsilon}\underline{\eta}^{-1}\max\limits_{\forall \phi\in\mathcal{P}}(1/\lambda_{\min}(P_{\phi}))}$, where $\tilde{\epsilon}$ and $\underline{\eta}$ are respectively given as in Theorem \ref{Thm_1} and Theorem \ref{thm_linear_MDSS} with ${\Upsilon}=\max(1,\max\limits_{\forall \phi,\hat{\phi}\in\mathcal{P}}\|\tilde{\Xi}_{\phi,\hat{\phi}}\|^2)$, $P_{\phi}$ satisfies \eqref{Lyapunov_eqn} with $\tilde{A}_{\phi}=I_{N_{\phi}-1}\otimes S-\varrho Z_{\phi}\otimes I_p$, $\forall \phi\in\mathcal{P}$.
Particularly,
 \eqref{open_MAS_comp} can reach (asymptotic) consensus if \eqref{open_MAS_delta_compact} is GUAS. }
\end{thm}
}
\begin{proof}  \upshape
Since \eqref{open_MAS_delta_compact} and \eqref{Agent_disagreement_moves} are respectively the special cases of \eqref{MDSS_linear_closed_loop} and \eqref{Mode_transform_var} with $A_{\sigma(t)}\triangleq I_{N_{\sigma(t)}-1}\otimes S$, $B_{\sigma(t)}K_{\sigma(t)}\triangleq -\varrho Z_{\sigma(t)}\otimes I_{p}$, $\Xi_{\sigma(t_k^+),\sigma(t_k^-)}\triangleq \tilde{\Xi}_{\sigma(t_k^+),\sigma(t_k^-)}$, $\Phi_k\triangleq \tilde{\Phi}_k$, $\check{\Xi}_{\sigma(t_k^+),\sigma(t_k^-)}\triangleq \check{\tilde{\Xi}}_{\sigma(t_k^+),\sigma(t_k^-)}$, then by Definition \ref{DEF_GUPS}, that \eqref{open_MAS_delta_compact} is GUPS (under state independent $\tilde{\Phi}_k$) implies $\limsup\limits_{t\rightarrow+\infty}\|\tilde{\delta}_{\sigma(t)}(t)\|\leq \epsilon$ for some non-negative constant $\epsilon$. It then follows that for any $i\in\{1,...,N_{\sigma(t)}-1\}$, $\limsup\limits_{t\rightarrow+\infty}\|\delta_{\sigma(t),i}(t)\|\leq \epsilon$, which, by $\delta_{\sigma(t),i}(t)\triangleq\xi_{\sigma(t),i}(t)-\xi_{\sigma(t),N_{\sigma(t)}}(t)$, leads to $\limsup\limits_ {t\rightarrow+\infty}\|\xi_{\sigma(t),i}(t)-\xi_{\sigma(t),j}(t)\|\leq{\varepsilon}$ with $\varepsilon=2\epsilon$, $\forall i,j\in\mathcal{V} _{\sigma(t)}$, i.e., \eqref{consensus_p} holds. This, by Definition \ref{consensus_GUPS}, indicates that the open MAS \eqref{open_MAS_comp} reaches the practical consensus. Moreover, by Theorem \ref{thm_linear_MDSS}, \eqref{open_MAS_delta_compact} and \eqref{Agent_disagreement_moves}, as well as the above result, the ultimate bound of the consensus error of the open MAS \eqref{open_MAS_comp} is then calculated by $\varepsilon=2\sqrt{\tilde{\epsilon}\underline{\eta}^{-1}\max\limits_{\forall \phi\in\mathcal{P}}(1/\lambda_{\min}(P_{\phi}))}$, where $\tilde{\epsilon}$ and $\underline{\eta}$ are respectively given as in Theorem \ref{Thm_1} and Theorem \ref{thm_linear_MDSS} with ${\Upsilon}=\max(1,\max\limits_{\forall \phi,\hat{\phi}\in\mathcal{P}}\|\tilde{\Xi}_{\phi,\hat{\phi}}\|^2)$, and $P_{\phi}$ satisfies \eqref{Lyapunov_eqn} with $\tilde{A}_{\phi}=I_{N_{\phi}-1}\otimes S-\varrho Z_{\phi}\otimes I_p$ for any $\phi\in\mathcal{P}$. On the other hand, given that \eqref{open_MAS_delta_compact} is GUAS (under state-dependent $\tilde{\Phi}_k$ satisfying \eqref{Agent_disagreement_moves_asymp}, which implies $\Upsilon=\max(1,\max\limits_{\forall \phi,\hat{\phi}\in\mathcal{P}}\|\check{\tilde{\Xi}}_{\phi,\hat{\phi}}\|^2)$), then similarly by Definition \ref{DEF_GUPS} we have $\lim\limits_{t\rightarrow+\infty}\|\tilde{\delta}_{\sigma(t)}(t)\|=0$, and for any $i\in\{1,...,N_{\sigma(t)}-1\}$, $\lim\limits_{t\rightarrow+\infty}\|\delta_{\sigma(t),i}(t)\|=0$, which indicates $\lim\limits_ {t\rightarrow+\infty}\|\xi_{\sigma(t),i}(t)-\xi_{\sigma(t),j}(t)\|=0$, $\forall i,j\in\mathcal{V} _{\sigma(t)}$, i.e., the asymptotic consensus is reached for \eqref{open_MAS_comp}. $\blacksquare$
\end{proof}

\begin{remark}\upshape
Theorem \ref{exo_consensus} indicates that
the consensus problem of the open MAS \eqref{open_MAS_comp} boils down to the stability problem of the corresponding $M^3D$ system \eqref{open_MAS_delta_compact} which can be readily handled with the aid of Theorem \ref{thm_linear_MDSS}. Note that the $M^3D$ system model \eqref{open_MAS_delta_compact} is in fact a special case of the linear $M^3D$ system \eqref{MDSS_linear_closed_loop}.
{Moreover, it is notable that for the considered open MAS \eqref{open_MAS_comp} to reach desired consensus performances, the TDADT conditions \eqref{TDADT_a} and \eqref{TDADT_b} are required to be satisfied for \eqref{open_MAS_delta_compact}. However, the calculations of both the lower and upper bounds of the TDADT in this case will require the eigenvalues of $I_{N_{\phi}-1}\otimes S-\varrho Z_{\phi}\otimes I_p$, $\phi\in\mathcal{P}$, which are typical global information due to the presence of $Z_{\phi}$. Similar cases also arise in other related works on switching topologies, see e.g., \cite[Eq. (12)]{RN648} and \cite[Eq. (33)]{RN921}. Nevertheless, this does not indicate the proposed method cannot be implemented in a distributed way. In fact, for implementation one does not  need to know the exact dwell-time bounds but only needs to make sure that these bounds are not violated. This means one can be more conservative when specifying a switching signal just to ensure the corresponding bounds are satisfied by a certain margin. Note also that the controller of each agent in \eqref{open_MAS} is still distributed (albeit not fully distributed).}
\end{remark}

\subsection{Simulation example}\label{section_4_4}
In this section, a simulation example will be presented to illustrate the above application.

Consider the open MAS \eqref{open_MAS_comp} with the following parameters:
$\varrho=3.75$, $S=[0.1~0.05;0~0.15]$, $L_1=[1	~ 0	 ~ 0 ~ -1; 0~  0 ~ 0 ~ 0;0	~ -1 ~	1 ~	0;
0 ~	0 ~	-1 ~ 1]$, $L_2=[0~ 0 ~   0;
   0  ~  0 ~   0;
    0 ~ -1 ~ 1]$, $L_3=[ 0 ~    0   ~  0 ~    0  ~ 0;
     0 ~    0 ~   0  ~  0 ~   0;
     0  ~  0  ~ 1 ~    0  ~  -1;
          0  ~   0    ~ 0   ~  0   ~  0;
     0 ~   0    ~ 0  ~  0  ~   0]$, $L_4=[1  ~ -1 ~   0;    0 ~  1 ~  -1;    0  ~ 0 ~   0]$.
The topologies are depicted in Fig. \ref{G_T}, where one can see that $\mathcal{G}_1$, $\mathcal{G}_4$ contain a directed spanning tree while $\mathcal{G}_2$, $\mathcal{G}_3$ do not. Then, by Lemma \ref{MAS_tra}, one derives for \eqref{open_MAS_delta_compact} that $Z_{1}=[1	~ 0	 ~ 1; 0~  0 ~ 1;0	~ -1~	2]$, $Z_{2}=[0	~ 1	; 0~  1]$, $Z_{3}=[0 ~ 0 ~0 ~0; 0~  0 ~ 0 ~0; 0~ 0~ 1~ 0;0 ~ 0 ~ 0 ~ 0]$, $Z_{4}=[1	~ -1; 0 ~ 1]$.
With Proposition \ref{MAS_tra2}, one readily concludes that $I_{N_{\phi}-1}\otimes S-\varrho Z_{\phi}\otimes I_p$ is Hurwitz for $\phi=1,4$ and non-Hurwitz for $\phi=2,3$.
Further, given $\eta_0=1$, $\chi_{1,2}=0.410$, $\chi_{1,3}=0.560$, $\chi_{2,1}=0.550$, $\chi_{3,1}=0.637$, $\chi_{3,4}=0.642$, $\chi_{4,2}=0.580$,
and applying Procedure \ref{Proc_1} to \eqref{open_MAS_delta_compact}, then the corresponding lower and upper bounds (denoted by $\tilde{\tau}_{i,j}$) of the (piecewise) TDADT $\tau_{i,j}(t_k,t_f)$, $k\in\{0,...,N(t_0,t_f)\}$, $t_0=0$, $t_f=12$ are derived as $\tilde{\tau}_{1,2}=2.053$, $\tilde{\tau}_{1,3}=1.901$, $\tilde{\tau}_{2,1}=0.429$, $\tilde{\tau}_{3,1}=0.323$, $\tilde{\tau}_{3,4}=0.317$, $\tilde{\tau}_{4,2}=1.522$.

\begin{figure}[!t]
\centering
\includegraphics[width=3.355in]{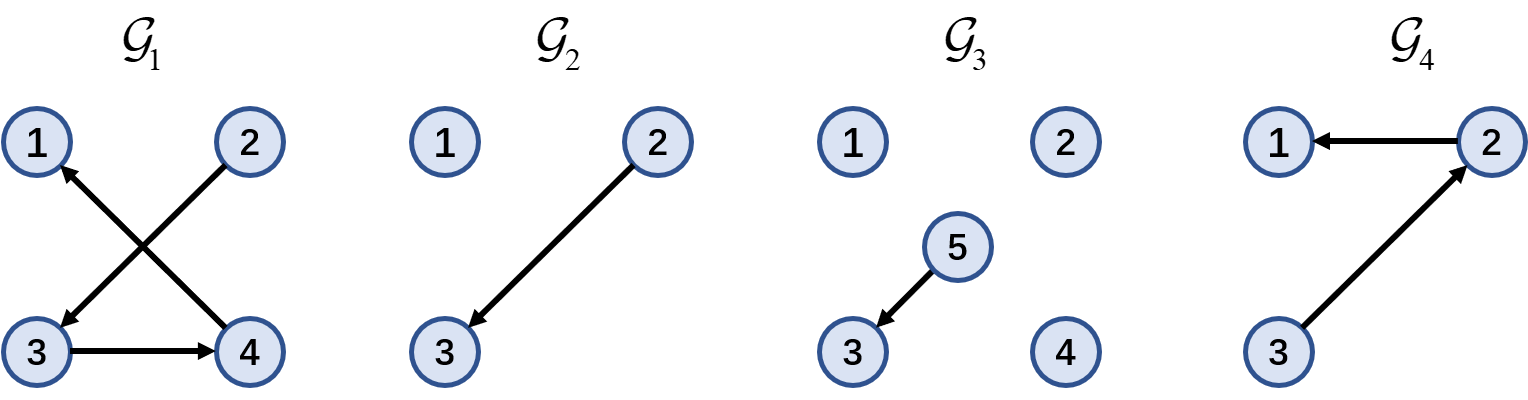}
\caption{Digraphs of the open MAS network considered in Section \ref{section_4_4}. The labeled circles
denote the nodes (agents) and the solid arrows denote the directed edges (connections among agents). It
is assumed in Section \ref{section_4_4} that only the agent with a larger label departs from
the group; newly incoming agents are labeled in sequence after the existing
largest label. Note that $\mathcal{G}_1$, $\mathcal{G}_4$ contain a directed spanning
tree while $\mathcal{G}_2$, $\mathcal{G}_3$ do not.}\label{G_T}
\end{figure}

\begin{figure}
\centering
\includegraphics[width=3.355in]{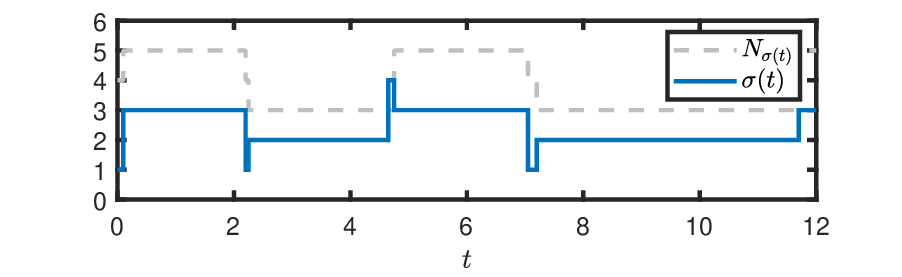}
\caption
{Switching signal $\sigma(t)$ (solid line) not satisfying \eqref{TDADT_a} and \eqref{TDADT_b}; evolution of the agent number $N_{\sigma(t)}$ (dash line).}
\label{Waveform_switching_counter}
\end{figure}
\begin{figure}
\centering
\includegraphics[width=3.355in]{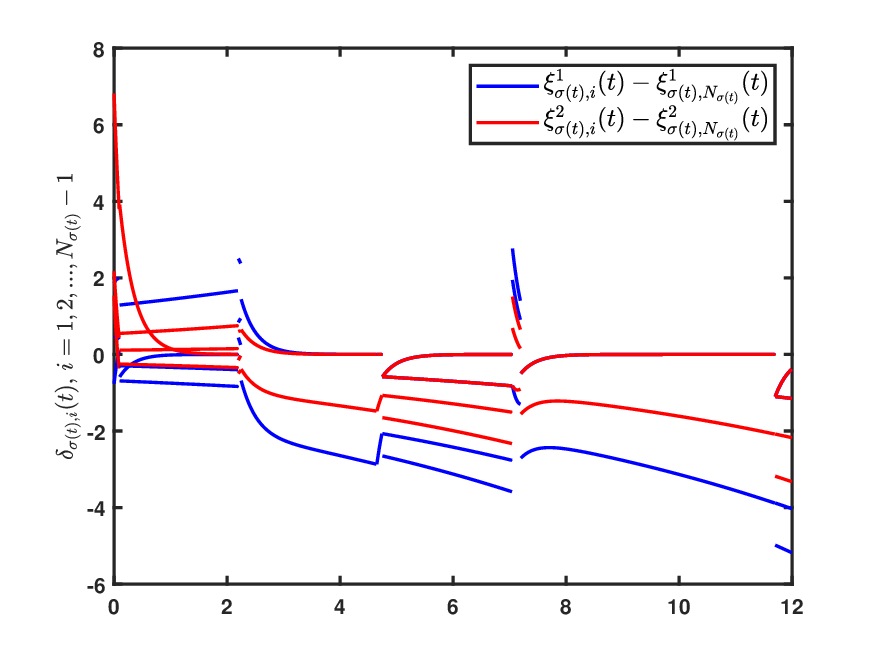}
\caption
{Consensus errors $\delta_{\sigma(t),i}(t)=\xi_{\sigma(t),i}(t)-\xi_{\sigma(t),N_{\sigma(t)}}(t)$, $i=1,...,N_{\sigma(t)}-1$ under $\sigma(t)$ of Fig. \ref{Waveform_switching_counter}.}
\label{State_disagreement_counter}
\end{figure}

\begin{figure}[!t]
\centering
\includegraphics[width=3.355in]{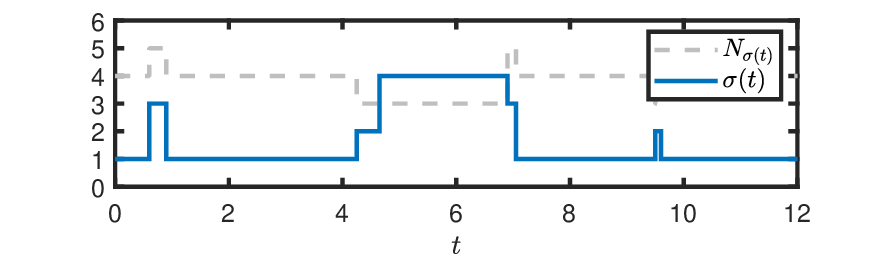}
\caption{Switching signal $\sigma(t)\in\tilde{\Psi}_{\sigma}$ (solid line) satisfying \eqref{TDADT_a} and \eqref{TDADT_b}; evolution of the agent number $N_ {\sigma(t)}$ (dash line).}
\label{Waveform_switching}
\end{figure}
\begin{figure}
\centering
\includegraphics[width=3.355in]{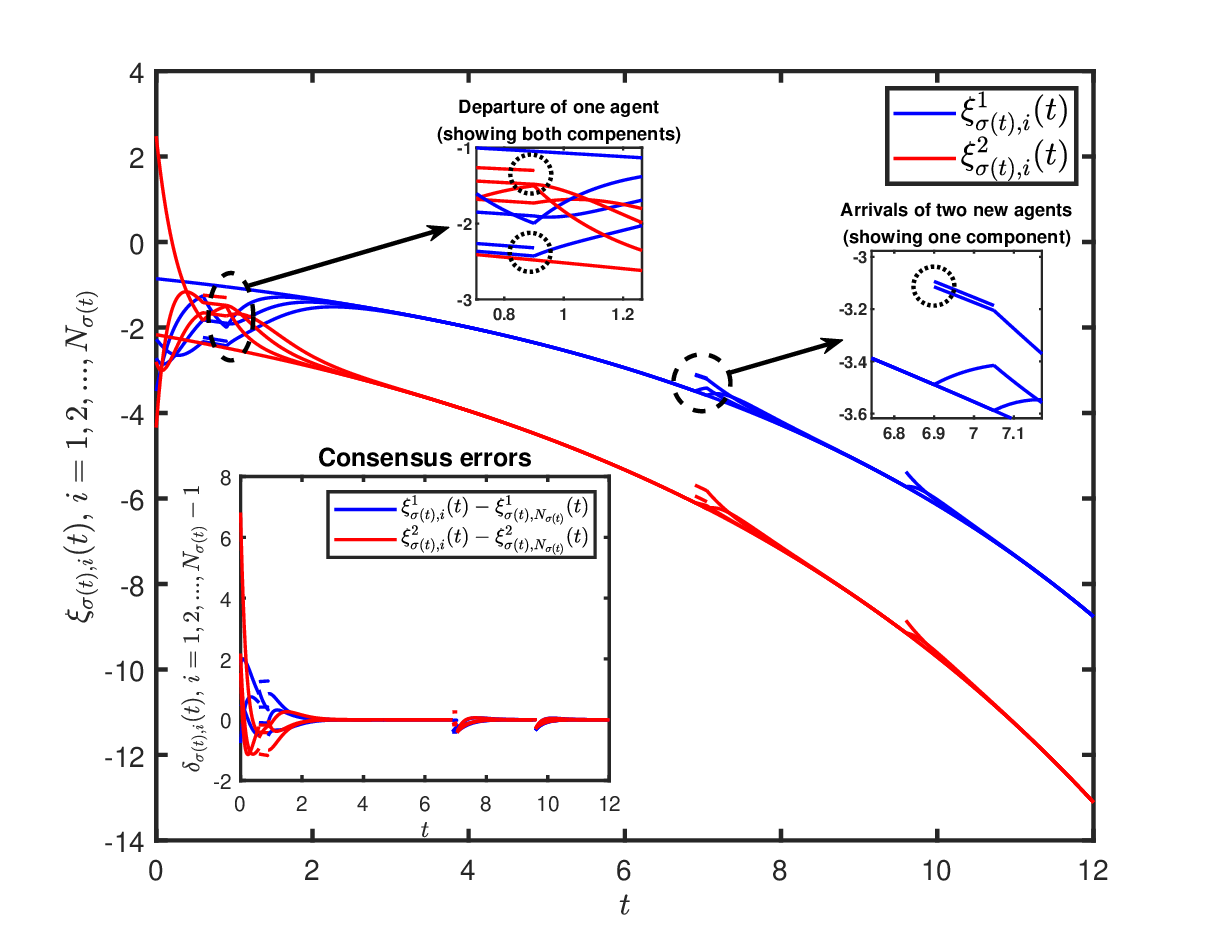}
\caption{Agent states $\xi_{\sigma(t),i}(t)$, $i=1,...,N_{\sigma(t)}$ under $\sigma(t)$ of Fig. \ref{Waveform_switching} and state-independent $\tilde{\Phi}_k$. The lower left subfigure depicts the corresponding consensus errors $\delta_{\sigma(t),i}(t)=\xi_{\sigma(t),i}(t)-\xi_{\sigma(t),N_{\sigma(t)}}(t)$, $i=1,...,N_{\sigma(t)}-1$. The black dash circles indicate two typical migration behaviors.}
\label{State_trajectories_impulse_not_convergent}
\end{figure}
\begin{figure}
\centering
\includegraphics[width=3.355in]{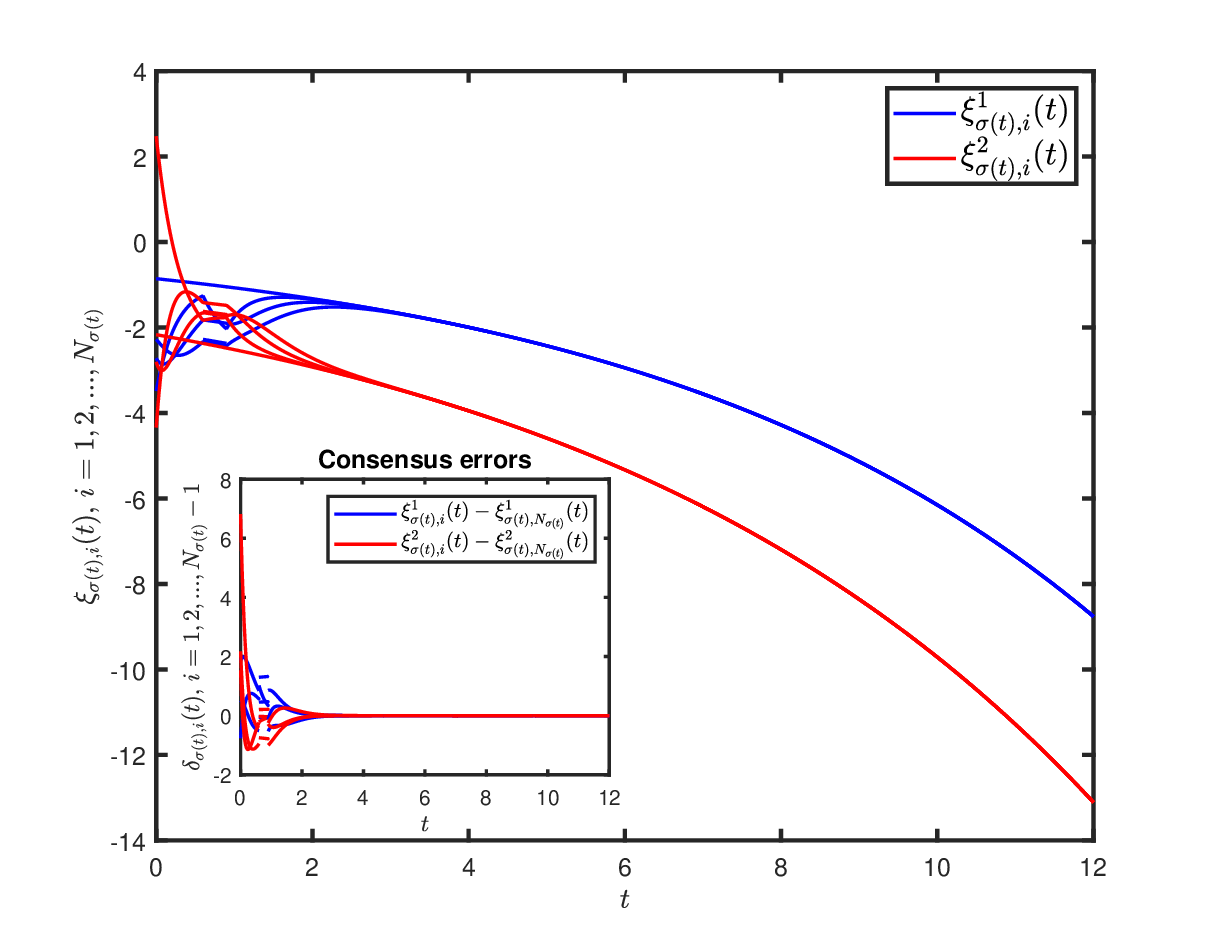}
\caption{Agent states $\xi_{\sigma(t),i}(t)$, $i=1,...,N_{\sigma(t)}$ under $\sigma(t)$ of Fig. \ref{Waveform_switching} and state-dependent $\tilde{\Phi}_k$ satisfying \eqref{Agent_disagreement_moves_asymp}. The  subfigure depicts the corresponding consensus errors $\delta_{\sigma(t),i}(t)=\xi_{\sigma(t),i}(t)-\xi_{\sigma(t),N_{\sigma(t)}}(t)$, $i=1,...,N_{\sigma(t)}-1$.}
\label{State_trajectories}
\end{figure}

Given the above parameter settings for the open MAS \eqref{open_MAS_comp}, we first consider a case where the switching signal $\sigma(t)$ does not satisfy \eqref{TDADT_a} and \eqref{TDADT_b} when $\tilde{\Phi}_k$ is state-dependent. The switching signal is plotted in Fig. \ref{Waveform_switching_counter} by a solid line and the evolution of the agent number is shown by a dash line. In this case, one can see from Fig. \ref{State_disagreement_counter} that despite under a state-dependent $\tilde{\Phi}_k$, the consensus errors ${\delta}_{\sigma(t),i}(t)$, $i=1,2,...,N_{\sigma(t)}-1$
still diverge with time, which implies that the considered open MAS cannot reach consensus under the switching signal given in Fig. \ref{Waveform_switching_counter}.

In contrast, we consider the same open MAS model with the switching signal $\sigma(t)$ depicted in Fig. \ref{Waveform_switching}, where the signal waveform is shown as a solid line and the  evolution of the number of agents $N_{\sigma(t)}$ is shown as a dash line. It can be readily seen from Fig. \ref{Waveform_switching} that the given switching signal satisfies $\sigma(t)\in\tilde{\Psi}_{\sigma}$ as well as the TDADT conditions \eqref{TDADT_a} and \eqref{TDADT_b} with the bounds derived above. First, consider state-independent $\tilde{\Phi}_k$. For simplicity, it is assumed in this example that the state impulse of $\tilde{\Phi}_k$ is only brought by the arrival of agents. The state values of the new arrived agents are randomly generated.
  The resultant trajectories of agent states $\xi_{\sigma(t),i}(t)$, $i=1,2,...,N_{\sigma(t)}$ and the consensus errors $\delta_{\sigma(t),i}(t)$, $i=1,2,...,N_{\sigma(t)}-1$ are depicted in Fig. \ref{State_trajectories_impulse_not_convergent}.
  It can be observed that with the non-vanishing property of the impulse $\tilde{\Phi}_k$ at each $t_k$, the consensus errors of the open MAS are ultimately bounded instead of asymptotically converging to zero, which indicates that the practical consensus is achieved. Note that the two zoomed-in windows of Fig.  \ref{State_trajectories_impulse_not_convergent} depicts the two typical agent migration behaviors of arrival and departure, respectively.
On the other hand, consider state-dependent $\bar{\tilde{\Phi}}_k$, which by \eqref{Agent_disagreement_moves_asymp} indicates that $\tilde{\Phi}_k$ is also state-dependent, and randomly generate ${\tilde{\hat{\Xi}}}_{\phi,\hat{\phi}}$ for $\phi,\hat{\phi}\in\{1,...,4\}$.
 The resultant trajectories of agent states and consensus errors are depicted in Fig. \ref{State_trajectories}. It can be seen that under the switching signal in Fig. \ref{Waveform_switching} and state-dependent $\tilde{\Phi}_k$ which brings only vanishing impulses, the consensus errors approach zero as $t$ goes on. This indicates the asymptotic consensus is achieved.

\section{Conclusion}\label{section_5}
We have studied the stability of the $M^3D$ system with different subsystem dimensions. The state transition of the system at each switching instant has been formulated as an affine map to incorporate both the dimension varying and the state impulsive effects.
In the presence of unstable subsystems and non-vanishing impulses, we show that the GUPS/GUAS of the $M^3D$ system can be ensured under the proposed (slow and fast) piecewise TDADT switchings, given that a series of Lyapunov-like conditions are satisfied. The stability conditions have then been verified for the linear subsystem case by the proposed parametric MLFs. Further, we have applied the result on the $M^3D$ system stability to the open MAS which features a size-varying switching topology, and
 show that the practical (asymptotic) consensus of the open MAS with disconnected digraphs boils down to the GUPS (GUAS) of the corresponding $M^3D$ system with unstable subsystems.
{Future endeavors can be made on further reducing the restrictiveness of the results obtained for the $M^3D$ system, such that they can apply to and be verified in more general cases (e.g., fully unstable/nonlinear subsystem dynamics).
Meanwhile, the future focus can also be put on real-world open MASs, such as the vehicle platoons with lane change maneuvers. Besides, experimental validations can also be considered in the place of simulations.}

\section*{Acknowledgment}

The authors would like to thank the editors and the
anonymous reviewers for their insightful comments and constructive suggestions on improving this work.


\bibliographystyle{plain1}

\bibliography{ref1}

\end{document}